\documentclass[11pt]{article}
\usepackage{bm}
\usepackage{amsmath}
\usepackage{amssymb}
\usepackage{fullpage}
\usepackage{graphicx}
\usepackage{hyperref}% add hypertext capabilities
\usepackage{cite}
\usepackage{authblk}
\usepackage{color}
\usepackage{lineno}

\newcommand{\red}{\textcolor{black}}

\PassOptionsToPackage{hyphens}{url}\usepackage{hyperref}

\title{\textbf{Perceptual Rationality: An Evolutionary Game Theory of Perceptually Rational Decision-Making}}
\author[1,2,3]{Mohammad Salahshour\thanks{salahshour.mohammad@gmail.com}}
\affil[1]{Department of Collective Behaviour, Max Planck Institute of Animal Behavior,
	78464 Konstanz, Germany.}
\affil[2]{Centre for the Advanced Study of Collective Behaviour, University of Konstanz, 78464 Konstanz, Germany.}
\affil[3]{Department of Biology, University of Konstanz, 78464 Konstanz, Germany.}

\begin{document}
	
	\maketitle

	\begin{abstract}
Understanding how biological organisms make decisions is of fundamental importance in understanding behavior. Such an understanding within evolutionary game theory so far has been sought by appealing to bounded rationality. Here, we present a perceptual rationality framework in the context of group cooperative interactions, where individuals make rational decisions based on their evolvable perception of the environment. We show that a simple public goods game accounts for power law distributed perceptual diversity. Incorporating the evolution of social information use into the framework reveals that rational decision-making is a natural root of the evolution of consistent personality differences and power-law distributed behavioral diversity. The behavioral diversity, core to the perceptual rationality approach, can lead to ever-shifting polymorphism or cyclic dynamics, through which different rational personality types coexist and engage in mutualistic, complementary, or competitive and exploitative relationships. This polymorphism can lead to non-monotonic evolution as external environmental conditions change. The framework provides predictions consistent with some large-scale eco-evolutionary patterns and illustrates how the evolution of social structure can modify large-scale eco-evolutionary patterns. Furthermore, consistent with most empirical evidence and in contrast to most theoretical predictions, our work suggests diversity is often detrimental to public good provision, especially in strong social dilemmas.
\end{abstract}

%\linenumbers

% Use "Eq" instead of "Equation" for equation citations.
\section{Introduction}
A focus on behavior has led many efforts to achieve a scientific understanding of life. In evolutionary game theory, this is exhibited in reliance on bounded rationality, where agents use predefined behavioral rules, such as simple heuristics or rules of thumb to make decisions \cite{Simon1955,Langlois1990,Szabo2007}. The dominance of bounded rationality in evolutionary game theory \cite{Szabo2007} has been fuelled by the limitations of pure rationality in accounting for empirical evidence \cite{Simon1955}. In game theory, pure rationality is defined based on payoff maximization, leading to the notion of Nash equilibrium \cite{Nash1950}. In the Nash equilibrium of a game, no player can increase their payoff by unilaterally changing their strategy. While such a notion of rationality is widespread in many disciplines, such as economics \cite{Gintis2000}, the limitations of this framework to account for behavior are evident \cite{Simon1955,Gintis2000,Fehr1999}. For instance, when such a notion is applied to social dilemmas, where individual and group interests conflict, a tragedy of the commons, resulting when individuals fail to cooperate, is inevitable \cite{Hardin1968}. Besides, such purely rational individuals, by definition, do not exhibit behavioral diversity. By contrast, behavioral diversity is a common characteristic of human \cite{Simon1955,Langlois1990} and animal \cite{Sih2012,Reale2007,Sih2004} decision-making.

These shortcomings have led to the introduction of bounded rationality as a more feasible paradigm for understanding behavior in the mid-20th century \cite{Simon1955}. In contrast to the logical structure of purely rational agents, boundedly rational agents rely on simple rules or heuristics to make decisions. This approach, which highlights the limits on cognitive processing, aligns with empirical observations that individuals do not always act in strictly rational ways and has proved instrumental across fields such as economics \cite{Langlois1990,Hanson1999}, psychology \cite{Kantor1963,Araiba2020}, and evolutionary game theory \cite{Szabo2007}. 

While bounded rationality has its merits, here, we argue that pure and bounded rationality represent two extremes; There is an alternative, which may be a more realistic representation of decision-making by living organisms in many cases. To do so, while retaining the core of pure rationality, that is payoff maximization based on logical thinking, we relax the assumption that individuals have immediate access to the objective reality (payoffs). This parallels some game theoretic frameworks, such as games with incomplete information \cite{Aumann1995}, Bayesian games \cite{Harsanyi1967}, and indirect evolutionary approach \cite{Guth1992,Heifetz2007,Guth1998}, where rational players face uncertainties or have evolvable preferences which can affect their decisions. We use this approach, to study how perceptions of the environment, together with other possible traits, such as social information use, may co-evolve with rationality, and shape behavior in rational agents.

In our framework, individuals are depicted as perceptual beings who gather information from the outside world, form perceptions, and make \textit{rational} decisions based on their perception. This perspective is in keeping with a host of empirical evidence indicating biological organisms are not solely rule-following agents---as depicted by bounded rationality, nor are they self-interested logical agents with immediate access to objective reality--- as depicted by pure rationality; They are perceptual beings who actively engage with their environments, gather information, form perceptions, and adjust their behavior based on these perceptions \cite{Eliassen2016,Polania2024}. Perceptions are shaped by evolutionary history and influenced by factors such as individual learning, social context, and experience, leading to an individualized and potentially biased view of the world, which organisms act upon. 

Although perception is shown to be a significant factor influencing both individual and group behavior in economic games \cite{Ellingsen2012,Deutsch1958,Eiser1974,Liberman2004,Sagiv2011,Columbus2020} and in shaping subjective utility functions \cite{Chen2024,Aknin2013,Nelson2016,Dakin2022}, the role of perception remains largely underexplored not only in evolutionary game theory but also in many fields of behavioral sciences. In contrast to this typical neglect, it is reasonable to expect that accounting for the perceptual nature of human and animal decision-making can advance our understanding of individual and collective behavior. The introduction of this perspective, and revealing some of its consequences for evolutionary game theory, is the purpose of this article. To do so, here, we appeal to public goods game as one of the most intensively studied paradigms for the evolution of cooperation in evolutionary games \cite{Szabo2007,Newton2018,Traulsen2023,Adami2016}. We reveal that accounting for individuals evolvable perception in such a simple environment can provide important insights on the evolution of behavior.

We begin with \nameref{Theory} by introducing a general framework of game theoretical interactions among rational agents with subjective assessment of their payoffs. In the bounded rationality framework, commonly appealed to in evolutionary game theory, the notion of rationality is overally dismissed and agents are defined by pre-defined behavioral rules. Such behavioral rules, while often fairly simple, such as cooperation and defection in public goods game \cite{Szabo2007}, can also be rather complex, such as in more sophisticated conditional strategies, in direct \cite{Hilbe2018,Schmid2021} or indirect reciprocity \cite{Riolo2001,Jansen2006}. By contrast, in our framework agents are not defined based on their behavior. Rather, individuals are rational and always play the Nash equilibrium of their game. Importantly, however, instead of having access to the objective reality, individuals' perceptions (as well as other possible traits, controlling their behavior) are subject to evolution. This makes individuals' behavior subject to evolution only indirectly. Such a perspective has been employed to study the evolution of preferences in the so-called indirect evolutionary approach \cite{Newton2018,Alger2019,Konigstein2000,Samuelson2001,Guth1992}. These works have argued for the evolutionary instability of objective rationality in binary interactions (two-person games) \cite{Heifetz2007}, leading to the evolution of cooperation in two-person two-strategy games \cite{Guth1998,Huck1999,Dekel2007}. 

After formulating a perceptual rationality framework in a general N-person, n-strategy game, we apply the framework to the evolution of cooperation in public good games both in the presence and absence of social information use. In the \nameref{Results} Section, we show that the framework can provide essential insights into a host of phenomena, ranging from the evolution of cooperation and behavioral diversity to consistent personality and social information use and even large-scale eco-evolutionary patterns. While it has been suggested that bounded rationality is essential to explain behavioral diversity manifest in the deviation of real-world decision-makers from the premises of rationality \cite{Simon1955,Langlois1990,Hanson1999}, the approach reveals that rational decision-making can be the very natural road to the evolution of behavioral diversity and consistent personality differences, that not only had been overlooked, but had been framed as the opposite. As we will detail in the \nameref{Discussion}, some of the concepts put forward in our framework have been largely unexplored or only poorly explored in evolutionary game theory. These include the evolution of consistent rational personalities, power law distributed behavioral, social, and perceptual diversity, the evolution of social information use, non-monotonic evolution of cooperative or social behavior as a function of external, environmental conditions due to shifting population configuration, or a connection between evolutionary game theory and large-scale eco-evolutionary patterns. Others, such as conditional cooperation, have been widely explored before. However, as we will argue, even in such cases, the new perspective employed here provides insights that, although parallel previous findings, are only conceptually related to previous work in evolutionary game theory.

\section{The Theoretical Framework}

\label{Theory}

\subsection{The Conceptual Framework}

We consider an $N$-person, $n$-strategy game. For simplicity of notation, we limit ourselves to symmetric games. However, in our framework, there is no essential difference between symmetric and asymmetric games, as even a symmetric game is typically perceived asymmetrically due to perspective-dependence (individual-dependence) of the reality (payoffs).

Each individual $i$ can play one of $n$ possible strategies, $(\sigma^1,..,\sigma^n)$. Individuals receive payoffs based on the strategies chosen by all the individuals, such that the payoff to individual $i$ is $\pi_i(\sigma_i, \sigma_{-i})$. Here, $\sigma_i$ is the strategy of individual $i$, and $\sigma_{-i}=\{\sigma_j|j\neq i\}$ is the strategy chosen by all the other individuals. Following standard game theory, we define rationality as maximizing own payoff \cite{Nash1950}. Thus, a rational individual plays the strategy $\sigma_i$, which maximizes its payoff, given the strategies chosen by all others, $\sigma_{-i}$. This notion of rationality leads to Nash equilibrium as the \textit{solution} of the game. The Nash equilibrium is defined as the strategy set, $(\sigma_i^*,\sigma_{-i}^*)$ where no player can increase its payoff by unilaterally changing its strategy. That is, in the Nash equilibrium:
\begin{align}
	\pi_i(\sigma_i^*,\sigma_{-i}^*)\geq \pi_i(\sigma_i,\sigma_{-i}^*) \quad\textit{for all i and for all } \sigma_i\neq \sigma_i^*.
\end{align} 

While retaining the notion of Nash equilibrium, we relax the assumption that individuals have immediate access to objective payoffs. Rather, our agents have beliefs or perceptions about the payoffs. In this regard, our framework has similarities and parallels with various game theoretic formulations where the assumption of perfect information about values of payoffs is relaxed. These include games with incomplete information \cite{Aumann1995}, or Bayesian games \cite{Harsanyi1967}, where, while players are rational, uncertainties about player types exist, indirect evolutionary approach \cite{Guth1992,Heifetz2007,Guth1998}, where rational players have evolvable preferences which can affect their decisions, and possibly stochastic games \cite{Shapley1953}, where environmental uncertainties can lead to objective uncertainties in game structure. 

Thus, in our framework, there is a distinction between actual payoff and perceived payoff. Each individual has a perceived payoff for each strategy configuration, $\hat{\pi}(\sigma_i,\sigma_{-i})$. Besides, the perceived payoff of different individuals may differ (e.g., some may value mutual cooperation more and some less). Consequently, while the objective game played by the individuals may be symmetric, the subjective game is, in principle always asymmetric. We parametrize the perceived payoffs by a set of trait values, $\bm{p}_i=\{p^1_i,..,p^m_i\}$. This set can represent traits, such as social values, cognitive biases, or personality traits, that individuals may acquire through evolution or learning and affect their perception of reality. Thus, the subjective payoff of the individual can be written as $\hat{\pi}_i(\sigma_i,\sigma_{-i};\bm{p}_i)$. 

As mentioned before, we define rationality in the same way as in classical game theory. However, instead of maximizing their objective payoffs, perceptually rational individuals maximize their payoffs over their subjective valuation of the payoffs by playing the Nash equilibrium of their perceived game:
\begin{align}
	\hat{\pi}_i(\sigma_i^*,\sigma_{-i}^*;\bm{p}_i)\geq \hat{\pi}_i(\sigma_i,\sigma_{-i}^*;\bm{p}_i) \quad\textit{for all i and for all } \sigma_i\neq \sigma_i^*.
\end{align} 

Therefore, while individuals are rational, in the sense of maximizing their payoffs given their perception of reality, their rationality is perspective-dependent. 

\subsection{Perceptual Public Goods Game}

We apply this general framework to the most extensively studied $N$-player game, a public goods game \cite{Szabo2007}. In this game, $g$ individuals play the game. Each individual has two strategies, cooperation and defection. Cooperators pay a unit cost to invest in the public good. Defectors pay no cost and do not invest. All the investments are multiplied by an enhancement factor, $r$, and are divided equally among the individuals. Thus, the net benefit accrued to a player from their investment is $r/g-1$, which is smaller than zero for $r<g$. This implies that the Nash equilibrium of this game is mutual defection, and individuals can increase their payoff by choosing to defect. However, if all individuals behave rationally, all receive a payoff of zero, which is lower than their payoff if they all cooperate (which is equal to $r-1$).

This argument has been used, for instance, by Hardin, to argue that ''freedom in the commons, brings ruin to all'' \cite{Hardin1968}. This conclusion, however, contrasts empirical observation according to which, while maintaining cooperation may be a challenge in many cases, cooperation is rather widespread \cite{Feeny1990,Ostrom1999}. Abandoning the notion of rationality altogether, previous work within evolutionary game theory widely employed the public goods framework to study the evolution of cooperation, suggesting diverse mechanisms, ranging from direct \cite{Trivers1971,Hilbe2018,Schmid2021} and indirect reciprocity \cite{Salahshour2022a,Nowak2005,Ohtsuki2006a,Boyd1989,Schmid2021,Nowak1998,Milinski2002,Kawakatsu2024} to punishment \cite{Salahshour2021C} and freedom of choice \cite{Keser2000,Hauert2002,Salahshour2021A,Salahshour2021B}, using boundedly rational agents.

To introduce the \textit{perceptual public goods game}, instead of boundedly rational or purely rational agents, we consider perceptually rational agents playing a public goods game (Fig. \ref{Fig1}\textbf{A}). Perceptually rational agents form perceptions of public good cost and benefit. Thus, we take $\bm{p}_i=(\hat{c}_i,\hat{b}_i)$, where $\hat{c}_i$ and $\hat{b}_i$ are the perceptions of individual $i$ of the personal cost of the public good and its benefit (which in reality is $r/g$). Without loss of generality, unless otherwise stated, we take $\hat{c}_i\in[0,1]$ and $\hat{b}_i\in[0,1]$. We have for the perceived payoff of such a perception, $\hat{\pi}_i(\sigma_i,\sigma_{-i})=\sum_j\delta_{\sigma_j,C}\hat{b}_i-\delta_{\sigma_i,C}\hat{c}_i=\delta_{\sigma_i,C}(\hat{b}_i-\hat{c}_i)+n_C\hat{b}_i$. Here, $n_C$ is the number of groupmates of the focal individual who cooperate, and $\delta_{\sigma_i,C}$ is a $\delta$ function which equals $1$ if $\sigma_i=C$ and zero otherwise.

Because $\hat{\pi}_i(C,\sigma_{-i})-\hat{\pi}_i(D,\sigma_{-i})=\hat{b}_i-\hat{c}_i$, for all $\sigma_{-i}$, a perceptually rational individual cooperates if $\hat{b}_i>\hat{c}_i$ and defects otherwise (for $\hat{b}_i-\hat{c}_i$=0 individual $i$ is indifferent, and both $C$ and $D$ are Nash equilibria). This argument shows that cooperation can be a Nash equilibrium of a public goods game among perceptually rational individuals if their perceived net personal benefit of the public good is positive. However, the key question is whether such distorted perceptions of reality can evolve in an evolutionary context.

\subsection{Social Perceptual Public Goods Game}
\begin{figure}%[ht]
	\centering
	\includegraphics[width=1\linewidth, trim = 50 0 35 24, clip,]{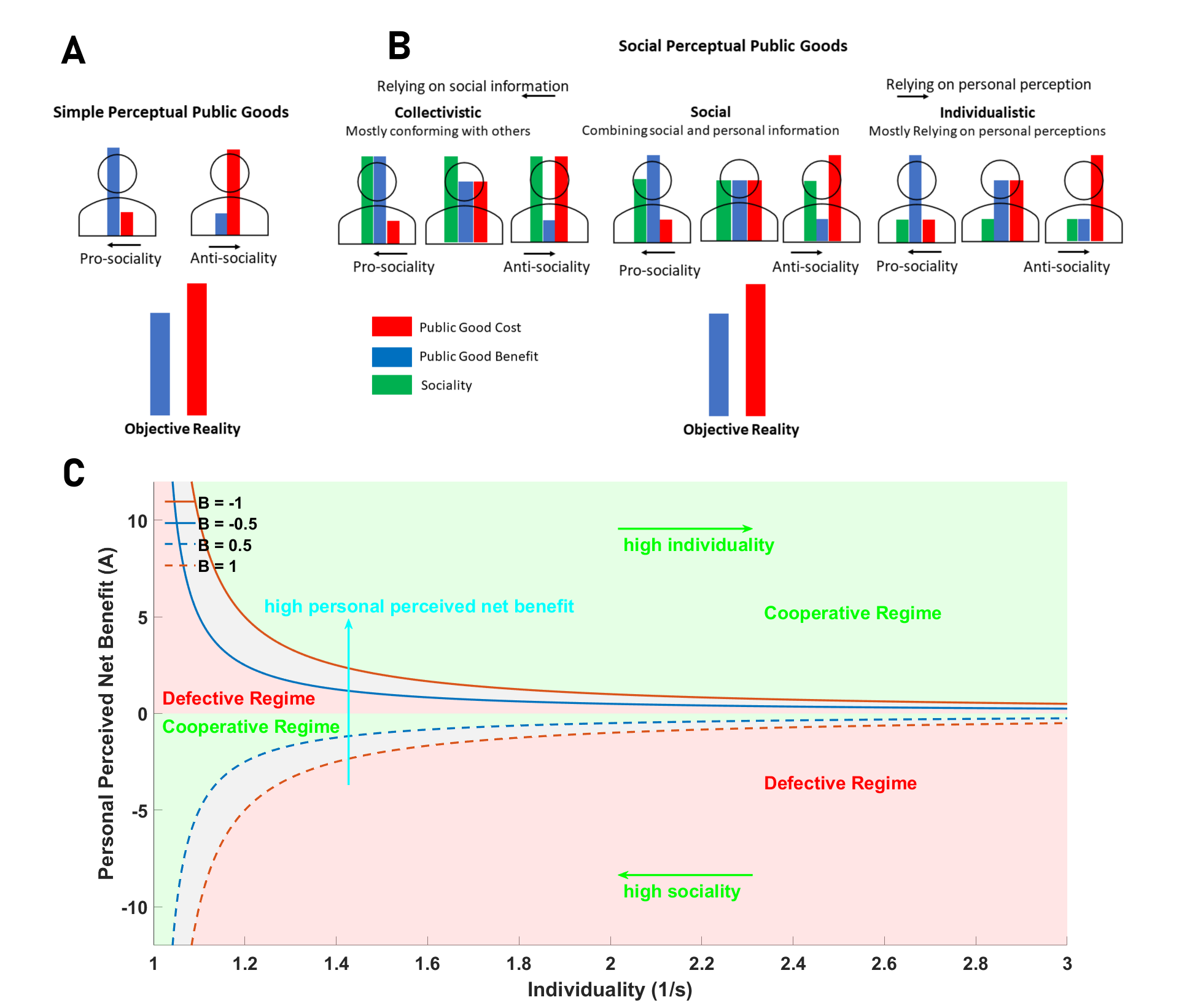}
	\caption{Perceptual public goods game. \textbf{A}: In a simple perceptual public goods game, rational individuals play a public goods game. Individuals are rational and play the Nash equilibrium of their game based on their perception of public good cost and benefit, which may or may not coincide with objective reality. \textbf{B}: In a social perceptual public goods game, in addition to their perception of public goods cost and benefit, individuals have a social trait that determines how susceptible they are to social information. Individuals play the Nash equilibrium of the game based on the socially perceived payoffs, defined as $\hat{p}^s_i=(1-s_i)\hat{p_i}+s_i\hat{p}_{-i}$, where, $\hat{p}_{-i}=\sum_{j\neq i}\hat{p}_j/(g-1)$, $i$ stands for the focal individual, $s$ is the sociality of the individual, $\hat{p}$ stands for the perception of individuals of the public good cost, $\hat{c}$, and benefit $\hat{b}$ and $g$ is the number of individuals playing the game. \textbf{C}: A rational individual cooperate if the difference between socially perceived payoffs of cooperation and defection, $(1-s_i)A+s_iB$, where $A=\hat{b}_i-\hat{c}_i$ and $B=\hat{b}_{-i}-\hat{c}_{-i}$, is positive and defect otherwise. Defining, the personal value of cooperation $\hat{\pi}_C=(1-s)A+sB$. Individuals can be decomposed into those with prosocial private economic perceptions, $A>0$ (green), and anti-social economic perceptions, $A<0$ (red). Increasing sociality may only lead to defection in the former, in groups with too low a perception of public goods benefit. By contrast, increasing sociality can only lead to cooperation in those with antisocial economic perception, when they are in groups with high economic perception. The value of sociality where individuals switch, $s^*$ depends on both the individual's perception, $A$, and their group's collective perception, $B$, leading to the flexibility of individuals' behavior. Besides, because others' perception depends on the focal individual's perception, individuals may develop complex strategies to adapt to, and affect, their group's behavior.}
	\label{Fig1}
\end{figure}

Because in our frameworks perceptions are the key traits shaping individuals' and collective's decision-making, the framework provides an approach to account for social learning contrasting the usual approach, where the focus is on behavior \cite{Molleman2013,Guzmán2007,Quan2022,Szolnoki2015,Szolnoki2012b,Yang2020} (see the \nameref{Discussion}). To see this, we consider a \textit{social perceptual public goods game}  (Fig. \ref{Fig1}\textbf{B}). This is a simple modification of a perceptual public goods game motivated by a wealth of empirical evidence that individuals within populations imitate others \cite{Evans2016}, or modify their beliefs based on social information, for instance, via social learning \cite{Laland2004,Kendal2018,Kendal2009} or conformity \cite{Sunstein2019}.

In this game, in addition to the perception of the public goods' cost and benefit, individuals have a social trait, which determines how susceptible they are to social information. Thus, we have, $\bm{p}_i=(\hat{c}_i,\hat{b}_i,s_i)$. Here, $\hat{c}_i$ and $\hat{b}_i$ are personal perceptions of public good cost and benefit, as before. $s_i\in[0,1]$ is a social learning trait, that determes how susceptible an individual is to social information. The socially learned value of public good cost and benefit for an individual is determined by both its private perceptions and social information: $\hat{b}^s_i=(1-s_i)\hat{b}_i+s_i\hat{b}_{-i}$, and $\hat{c}^s_i=(1-s_i)\hat{c}_i+s_i\hat{c}_{-i}$, where, $\hat{b}_{-i}=\sum_{j\neq i}\hat{b}_j/g$ and $\hat{c}_{-i}=\sum_{j\neq i}\hat{c}_j/g$, are the average private perception of others in the group, of the public good cost and benefit, respectively. Individuals play the Nash equilibrium of their perceived game, based on the socially learned payoffs, $\hat{b}^s$ and $\hat{c}^s$. For $s=0$, individuals are fully individualistic, ignoring social information, and for $s=1$, individuals are fully collectivistic, adopting the collective perception of others in their group.

We have for the difference between perceived payoffs of an individual when they cooperate and defect, $\hat{\pi}_i^s(C,\sigma_{-i})-\hat{\pi}_i^s(D,\sigma_{-i})=\hat{b}^s_i-\hat{c}^s_i=(1-s)(\hat{b}_i-\hat{c}_i)+s(\hat{b}_{-i}-\hat{c}_{-i})$ $= (1-s_i)A+s_iB$, where we have defined $A=\hat{b}_i-\hat{c}_i$ and $B=\hat{b}_{-i}-\hat{c}_{-i}$. Defining, the personal value of cooperation $\hat{\pi}_C=(1-s)A+sB$, the individual cooperates if $\hat{\pi}_C>0$ and defects otherwise. Now, in addition to its own perception, $A$, the behavior of an individual depends on both its sociality and its group's average perception. This introduces a complex decision-making where individuals' behavior is shaped by their perception, sociality, and group-level information. Group perception, and thus behavior, in turn, is shaped by the perception of the individuals. Out of this feedback between individual and group decision-making, different types of rational decision-making may emerge. To see this, we separate prosocial and antisocial (economic) perceptions based on the individual's private perceptions:
\begin{align}
	\begin{cases}
		\text{Prosocial Perception ($A>0$): }\hat{\pi}_C>0\Rightarrow (1/s-1)>-B/A,\\
		\text{Antisocial Perception ($A<0$): }\hat{\pi}_C>0\Rightarrow (1/s-1)<-B/A
	\end{cases}
\end{align}
An individual with a prosocial private perception, cooperates if it relies on their own perception (small $s$). However, when $s\neq 0$, the individual integrates the collective perception of its group into its own perception to make a decision. The individual switches to defection, when their sociality satisfies, $1/s=-B/A+1$. Hence, such an individual always cooperates in a group with an average positive perception of the public good net benefit. However, as the perception of others in the group decreases, the individual switches to defection, when $B$ reaches, $A(1/s-1)$. On the other hand, an individual with antisocial economic perception, $A<0$, defects for small $s$, and as $s$ increases, the individual may switch to cooperation provided the collective perception of the group is positive enough. In both cases, because the behavior of the individuals depend on others' perceptions, one can consider such individuals conditional cooperators. However, we have a continuum of such conditional cooperating individuals, who switch to defection in groups with an increasingly lower perception of public good net benefit, depending on their own perception, $A$, and and their sociality. Besides, the effect of sociality on behavior is different for the two categories of individuals. While for those with positive private perception ($A>0$) increasing sociality can only induce defection, for those with negative private perception, increasing sociality can only induce cooperation (see Fig. \ref{Fig1}\textbf{C}).

The private perception of the individual not only affects their own behavior but also affects the behavior of others in the group. Thus, individuals can employ different decision-making ``strategies''. We note that strategy here is used in a different sense than cooperation and defection. Instead, it refers to the specialization of individuals in developing different sociality and private perceptions, which may or may not perform well in a competitive context. As such, we refer to such decision-making strategies, as rationality types. Below, we discuss some possible rationalities:

$\bullet$ Collectivistic cooperators have high positive $A$ and high $s$ ($A\to1$ $s\to1$). These individuals copy the perception of others in their group. Thus, they cooperate in groups with positive perceptions and defect otherwise. Because they have a high private perception, they increase the likelihood of cooperation in their group. Nevertheless, due to having high sociality, they only marginally incorporate their own perception in their decision-making. As such, these individuals employ a deceptive strategy. Besides, such individuals show a \textit{contextual consistency} \cite{Kaiser2021} of personality, as they develop a prosocial personality in two different dimensions of behavior (economic perception and social information use).\\
$\bullet$Collectivsitic defectors have large negative $A$ and high $s$ ($A\to -1$ $s\to1$). Similarly to collectivistic cooperators, collectivistic defectors copy the perception of their group and thus cooperate in groups with high perception and defect otherwise. However, these individuals, having a negative economic perception, decrease the likelihood of cooperation in their group. Being prosocial in information use and anti-social in their economic perception, these individuals do not show contextual consistency.\\
$\bullet$ Individualistic cooperators ($A\in[0,1]$ $s\to 0$): These individuals develop positive $A$ and small $s$. Thus, they make individualistic decisions by not relying on their group perception. Such individuals can be considered unconditional cooperators, who do not exhibit contextual consistency. While do not employ a collective decision-making approach, they nevertheless can affect the group decision-making by affecting others' perceptions. Thus, these individuals can be refined into sub-categories, with $A\to 1$, who influence others in favor of cooperation (and can be called influential individualistic cooperators), and $A\to0^+$ who only marginally influences their group decision-making (socially neutral individualistic cooperators).\\
$\bullet$ Individualistic defectors ($A\in[-1,0]$ $s\to 0$): These individuals develop negative $A$ and small $s$ and are unconditional defectors with consistent antisocial perception and social information use. Such individuals can also be decomposed into finer categories, such as influential individualistic defectors ($A\to -1$), who influence the group in favor of their antisocial perspective, and socially neutral individualistic defectors ($A\to 0^-$), who only marginally affect the collective perception of the group. \\
$\bullet$ Economically neutral perceptions: When $A=0$, individuals are neutral in their private perceptions. For $s>0$, these individuals simply act based on the collective perception of their group. When $s=0$, however, these individuals randomize their strategy.\\

The above categories represent only a small portion, but rather extreme types, of rational decision-making strategies. Social individuals who develop average $s$, lie in between. Such individuals integrate personal and collective perceptions in making decisions. Depending on the exact value of $A$ and $s$, such individuals' behavior changes in groups with different perceptions, and they may affect the collective dynamics of the group toward cooperation and defection to varying degrees (Fig. \ref{Fig1}\textbf{C}). This enumeration of some possible types thus shows the complexity of rational decision-making. An important question is, however, which of such possibilities performs well in a competitive, evolutionary context.

\subsection{Lack of Evolutionary Stable Strategies and Behavioral Diversity}

To address this question, we consider a simplified version of the model, with binary perceptions and sociality. Specifically, each individual’s traits for the perception of public good cost (\(\hat{c}\)), benefit (\(\hat{b}\)), and sociality (\(s\)) can be \(0\) or \(1\). This gives a total of eight possible types. We denote these types by a sequence of three binary numbers, such as \((0,1,1)\), meaning private cost equal to $0$, private benefit $1$, and fully social. In this simplification of the model, the collective perception of the individual is defined as, $p_i^s=\sum_i p_i/g$. Here, the summation is over all the individuals in the group, including the focal individual and $p$ stands for $\hat{c}$ or $\hat{b}$. Explicitly, while in the model with continuous traits, social averaging is over other individuals in the group, in the binary model, social averaging is over all individuals including the focal individual. The reason for this choice is that in the model with a continuous trait, the individual can naturally integrate their own perception with others by freely weighting other perceptions by any number between $0$ and $1$. Given that this freedom is lost with binary sociality, it is more reasonable to allow for the integration of the focal individual's perception in social averaging by slightly changing the implementation.

We begin by showing that no type is evolutionary stable. In classical evolutionary game theory, a ``strategy'' (which, better to be referred to as ``type'' here) is said to be evolutionary stable if it is resistant to the introduction of mutants. Following Maynard Smith \cite{Smith1973}, a type $Y$ is said to be an evolutionary stable strategy (ESS) if the following two conditions are satisfied. Firstly, the payoff of the resident against itself, \(\Pi(Y\mid Y)\), should be larger than the payoff of any mutant $X$ against the resident \(\Pi(X\mid Y)\):
\[
\Pi(Y\mid Y) > \Pi(X\mid Y)
\quad\text{for all }X\neq Y,
\]
Secondly, if there is a tie for any mutant, $\Pi(Y\mid Y) = \Pi(X\mid Y)$, the resident should perform better in a population of all mutants:
\[
\Pi(Y\mid Y) \;=\;\Pi(X\mid Y)
\;\Longrightarrow\;
\Pi(Y\mid X) \;>\;\Pi(X\mid X).
\]
In the \red{Supplementary Note 1, S.2}, by systematically checking all eight binary types, we find that, in the public good regime \(1<r<g\), no single type satisfies these conditions. Although certain pairwise comparisons show that a resident type \(Y\) can successfully resist one specific mutant, there is always a mutant \(X\) that can succeed against \(Y\) by exploiting (or outcompeting) it under the second condition. Particularly, individualistic anito-social types, $(1,0,0)$ (who unconditionally defect) are resistant against prosocial types, $(0,1,.)$, (here, a dot means the statement is valid for both individualistic and social types). The key to the lack of ESS is social randomizing types, $(0,0,1)$ and $(1,1,1)$, who copy the strategy of others. While these types perform better against individualistic anito-social types, $(1,0,0)$, they perform worse against prosocial types, $(0,1,.)$, who, in turn, perform poorly in competition with anti-social types.

Thus, the lack of evolutionary stability is due to the fact that a single type's \emph{behavioral decision} (cooperate vs. defect) can still change depending on local context (the perception of others), since each agent is playing a rational response based on its perceived payoff. Thus, the payoff to any given type \((c,b,s)\) depends on the frequencies of other types in its group and how social averaging may shift each agent's perceived cost and benefit. Consequently, the selection gradients are highly configuration‐dependent. As a result, rational individuals end up specializing in complementary ``roles'' that make use of different combinations of \((c,b,s)\). As we will see, this leads to consistent personalities, where individuals with, e.g., a negative cost-benefit perception also tend to remain asocial (and so defect), whereas individuals who see a positive net benefit more often evolve to be social (and thus follow others' perceptions).

An important consequence of the lack of evolutionary stability is that the evolutionary dynamics exhibits complex dynamics where multiple types coexist in ever‐shifting polymorphisms exhibiting a ``rock--paper--scissors''--like dynamics. As we will see, such population heterogenity can occur both in time, where different types exhibit cyclic dynamics, or diverse types may coexist at the same time, depending on the group size, $g$.

\subsection{The Evolutionary Model}

To study the evolution of perceptually rational individuals we consider a standard evolutionary algorithm corresponding to a simple replicator dynamics \cite{Nowak2006b}. We consider a non-overlapping generation model in which individuals in each generation play the game, gather payoffs, and produce offspring proportional to their payoffs subject to mutations. The offspring compose the population pool in the next generation. 

Specifically, we consider both well-mixed and structured populations. In a well-mixed population, in each generation, groups of $g$ individuals are formed at random out of the population pool of $N$ individuals to play the game, such that the whole population is divided into randomly formed groups of $N$ individuals. In a structured population, we consider individuals residing on a first nearest-neighbor lattice with periodic boundaries and von Neumann connectivity (where each individual is connected to four neighbors to its right, left, top, and bottom). We choose a lattice, because this is a simple homogenous network employed in most of the past studies \cite{Szabo2007,Perc2013}, allowing us to ensure the phenomenology is not due to complex connectivity structure, and retaining comparability of our results with many past studies where the same network is employed. Following many past studies, in a structured population, a public good is staged centered on each individual and all four neighbors of the individual participate in the public goods. Thus, each individual participates in five public goods and their payoffs are their average payoffs across the five public goods. The exact number of public goods played by the individuals only acts as an averaging variable and thus does not affect the results.

In addition to receiving payoffs from the game, individuals receive a base payoff, $\pi_0=1$ from other activities not related to the public goods and reproduce with a probability proportional to their payoff. Offspring inherit the traits of their parent subject to mutations. Specifically, we consider several modifications of the model as follows:

\noindent$\bullet$ In the simple perceptual public goods games, these traits are the perception of the individuals of public good cost and benefit, $\hat{c}\in [0,1]$ and $\hat{b}\in[0,1]$, respectively. \\
\noindent$\bullet$ In the social perceptual public goods game, the evolvable traits are the perceptions of public goods cost and benefit, together with a social trait, $\hat{c},\hat{b},s\in[0,1]$. \\
\noindent$\bullet$ In the social perceptual public goods game with binary traits, the evolvable traits are the binary perceptions of public goods cost and benefit, together with a social trait, $\hat{c},\hat{b},s \in{0,1}$.\\
\noindent$\bullet$ We also consider two simplifications of the social perceptual public goods game where rational individuals have access to the objective value of the public good personal cost (which is always $1$) or benefit (which in reality is $r/g$), but their perception of, respectively, public good benefit and cost is subject to evolution. In this simplified model, to allow diverse behavior we limit the evolvable perception in the interval $[0,2]$.\\

Mutations in each of the traits occur independently, and each with probability $\nu$. In the case of mutation, the trait of the offspring is chosen uniformly at random in the admissible interval ($[0,1]$, ${0,1}$, or $[0,2]$, depending on the model variant). In the \red{Supplementary Information, S.6}, we consider a more complex mutation process where mutations can occur to a small neighborhood of the parent's trait and show that the exact way mutations are introduced in the model does not affect the results.

We analyze binary social public goods games using both the replicator dynamics, derived in the \nameref{Methods} Section and simulations and other model variants are analyzed using simulations. In most of our analysis, we take the group size, $g=5$. This choice is employed in most of the past theoretical and empirical studies \cite{Newton2018,Szabo2007,Perc2013}, partly because many studies have used a lattice (with von Neumann connectivity where each individual has four neighbors), where a group size of $5$ is dictated by the neighborhood structure. However, we study the effect of group size by investigating groups of $g=4$ and $g=3$ as well. Unless otherwise stated, all the simulations start with an initial condition which is the most difficult for the evolution of cooperation. That is, individuals have zero perception of benefit, and randomly chosen perceptions of public good cost. In the social perceptual public goods game, initially, all the individuals are fully individualistic and have $s=0$. See \red{S.1 for further details}.

\subsection{Interpretations of the evolutionary dynamics}

Before proceeding, it is worth discussing two points concerning the interpretations of the evolutionary dynamics. Firstly, the evolutionary dynamics employed here admit two different interpretations \cite{Nowak2006b}:\\
1. A non-overlapping generation model mimicking biological evolution. In this interpretation, evolvable traits are genetic traits that are fixed during the individual lifetime. Thus, in this interpretation, individuals' perceptions (and sociality) should be regarded as genetic traits that are fixed over their lifespan (up to social learning in the social perceptual public goods game). While not a realistic assumption in many cases where individuals' perceptions are subject to learning during their lifespan, this can be a reasonable assumption in some cases where perceptions represent cognitive biases that are stable over an individual lifespan.\\
2. The same evolutionary dynamics (replicator dynamics) can also be understood as a simple imitation dynamic in a social system \cite{Nowak2006b}. In this interpretation, evolution is thought of as social learning by the same individuals over successive stages. Thus, ``death'' is not understood as biological death, but rather as the adoption of a new ``strategy'' by the same individual over time (``generations''). In this interpretation of our evolutionary dynamics, the assumption that individuals' perceptions are fixed over their lifespan does not hold. Rather, the evolutionary dynamics accounts for how perceptions spread in the population pool when individuals act upon such perceptions and update their perceptions by imitating more successful individuals (which can be represented by, e.g., replicator dynamics \cite{Nowak2006b,Szabo2007}).

The second point concerns the flexibility of the behavior of rational decision-makers. As mentioned before, our agents are not defined by fixed behavioral rules, rather they are perceptually rational individuals who play the Nash equilibrium of their perceived game. This implies that the same individual acts differently in different contexts (i.e., in different groups), which we refer to as the flexibility of behavior of rational agents. Since we consider a simple non-overlapping generation model with one-shot interactions, each ``individual'' plays only one game (in a well-mixed population). Thus using the term flexibility of behavior may seem misleading. However, two points are in order. Firstly, it is intuitively clear that allowing individuals to play more than one game and in different groups (as we do in a structured population), does not change the outcome of evolutionary dynamics as the number of interactions is an averaging variable that does not change the \textit{expected payoffs} of the individuals. Thus, in the limit of infinite population size, all such evolutionary dynamics are described by the same replicator dynamics, irrespective of the number of games that individuals play. In a finite population, the number of interactions can affect the ``noise'' (deviation of the realized payoffs from the expected payoff), and thus, similar dynamics can be restored by increasing the population size to suppress the noise.

Secondly, as it typically happens in evolutionary dynamics like ours (i.e., replicator dynamics), as long as the mutation rate is below $1$, different copies of the same trait exist in the population. Such individuals with identical traits, all representing the same decision-maker, experience different environments (i.e., play their games in different groups), and act differently based on their group context. In this sense, such flexibility of behavior is indeed exhibited in our one-shot interaction model.

\section{Results}
\label{Results}

\subsection{Perceptual public goods game}

\begin{figure}%[ht]
	\centering
	\includegraphics[width=1\linewidth, trim = 0 0 5 4, clip,]{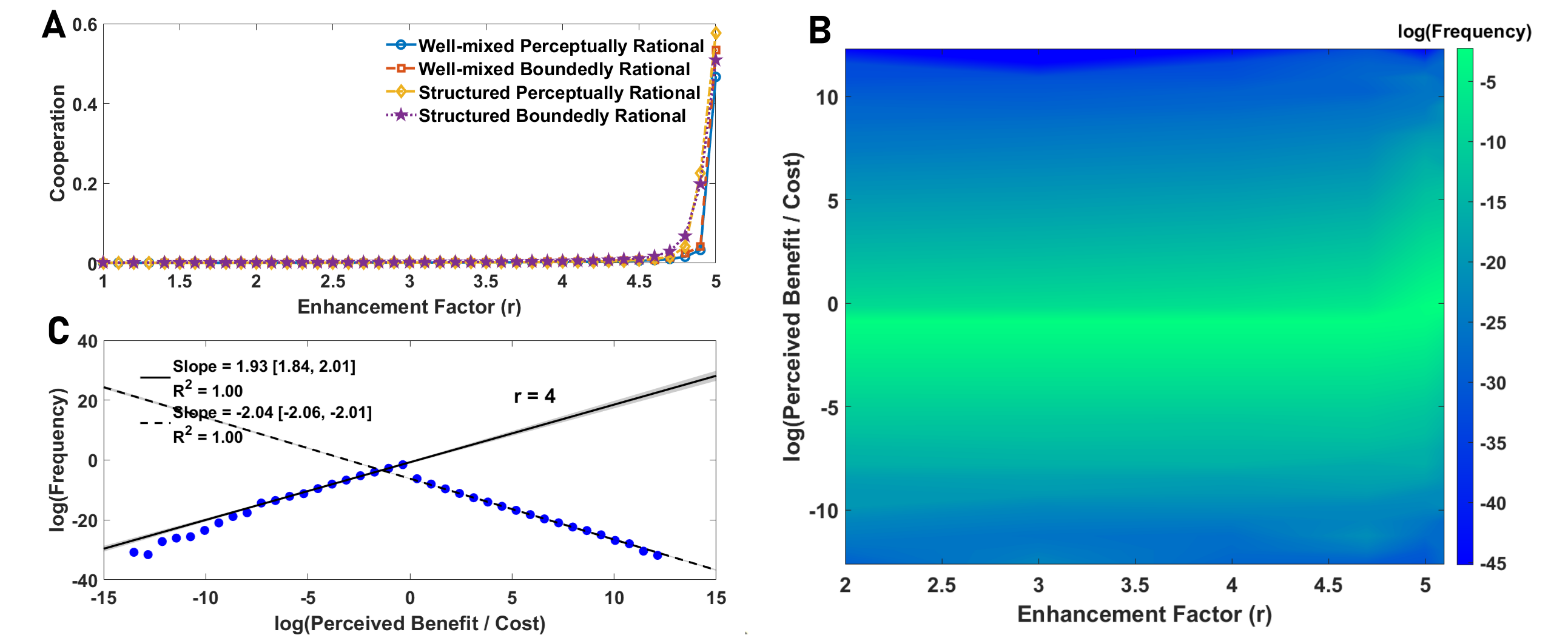}
	\caption{The evolution of perceptual diversity in a simple perceptual public goods game. \textbf{A}: Cooperation in a simple perceptual public goods game in both well-mixed and structured populations is compared with cooperation in boundedly rational agents. In both boundedly rational and perceptually rational agents, cooperation does not evolve (in the public good regime, $1<r<5$, where a social dilemma exists) in a simple public goods game. \textbf{B}: The logarithm of the frequency perceived benefit-to-cost ratio is plotted as a function of enhancement fact. Despite the lack of cooperation and behavioral simplicity, individuals exhibit a broad distribution of perceptions. \textbf{C}: The logarithm of the frequency of perceived benefit-to-cost ratio as a function of the logarithm of benefit-to-cost ratio is plotted. The perceived benefit-to-cost ratio exhibits a nearly symmetric power law with an exponent equal to $2$. Parameter values: $\nu=10^{-3}$ and $N=10000$.  }
	\label{Fig2}
\end{figure}

We begin with a simple perceptual public goods game. In Fig. \ref{Fig2}\textbf{A}, we plot the frequency of cooperation as a function of the enhancement factor in both well-mixed and structured populations. For comparison, we also present results for a simple public good game played by boundedly rational agents. A simplified version of the model where individuals can develop binary perceptions of the cost and benefit can be solved in terms of the replicator dynamics and is discussed in the \red{Supplementary Notes, S.3}. 

The results show that cooperation evolves only for relatively large enhancement factors and is largely similar for boundedly rational and perceptually rational agents. The similarity of the evolutionary dynamics of boundedly rational and perceptually rational agents results from the fact that a perceptually rational agent cooperates if their perceived benefit of the public good is larger than its perceived cost, $b>c$, and defects otherwise, leading to the simplicity of their behavior (\red{see S. 3}). 

Despite its behavioral simplicity, a simple perceptual public goods game gives rise to power-law distributed diversity of perceptions. To see this, in Fig. \ref{Fig2}\textbf{B}, we color plot the distribution of the logarithm of the perceived benefit-to-cost ratio as a function of the enhancement factor. While, for small enhancement factors, most individuals evolve a higher perceived cost than perceived benefit, preventing them from cooperating, a broad tail for this distribution is observed. In many complex systems, diversity is associated with power laws \cite{Newman2005,Stanley2000,Brown2002}, which can lead to maximum diversity \cite{Marsili2022,Marsili2013,Corominas2018,Newman2005,Brown2002}. In Fig. \ref{Fig2}\textbf{C}, we present the distribution of the logarithm of the frequency of the perceived benefit-to-cost ratio. The results indicate a nearly symmetric tent-shaped power law with an exponent close to $2$. Similar power laws are detected for the tail of the inverse cost and benefit, often with an exponent close to $-2$, which implies the power-law distribution of the tail of cost and benefit with a small exponent (\red{see S. 4}). This fact shows that while strong selection acts on behavior, there is no or only a weak selection against perceptual diversity, and individuals with diverse perceptions of public good cost and benefit evolve.

\subsection{Social perceptual public goods game}

A simple perceptual public goods game falls short of promoting cooperation and behavioral diversity. However, it can provide a foundation upon which an evolutionary dynamic with a rich phenomenology can be built. This is true because there is no reason why increasing the complexity of the evolutionary dynamics should remove the diversity inherent in the evolutionary dynamics of perceptually rational agents. In contrast, intuition suggests considering more realistic contexts can only increase perceptual diversity and lead to behavioral diversity.

To see this, we begin with a simple observation: many biological organisms are social, exhibit social learning, and their perceptions are influenced by others \cite{Evans2016,Laland2004,Kendal2018,Kendal2009}. This contrasts a simple perceptual public good game, where individuals can only develop private perceptions. How does such social susceptibility to others' perceptions, or in other words, the evolution of collective perception of the environment affect behavioral dynamics?

\begin{figure}[!hbt]
	\centering
	\includegraphics[width=1\linewidth, trim = 0 4 5 4, clip,]{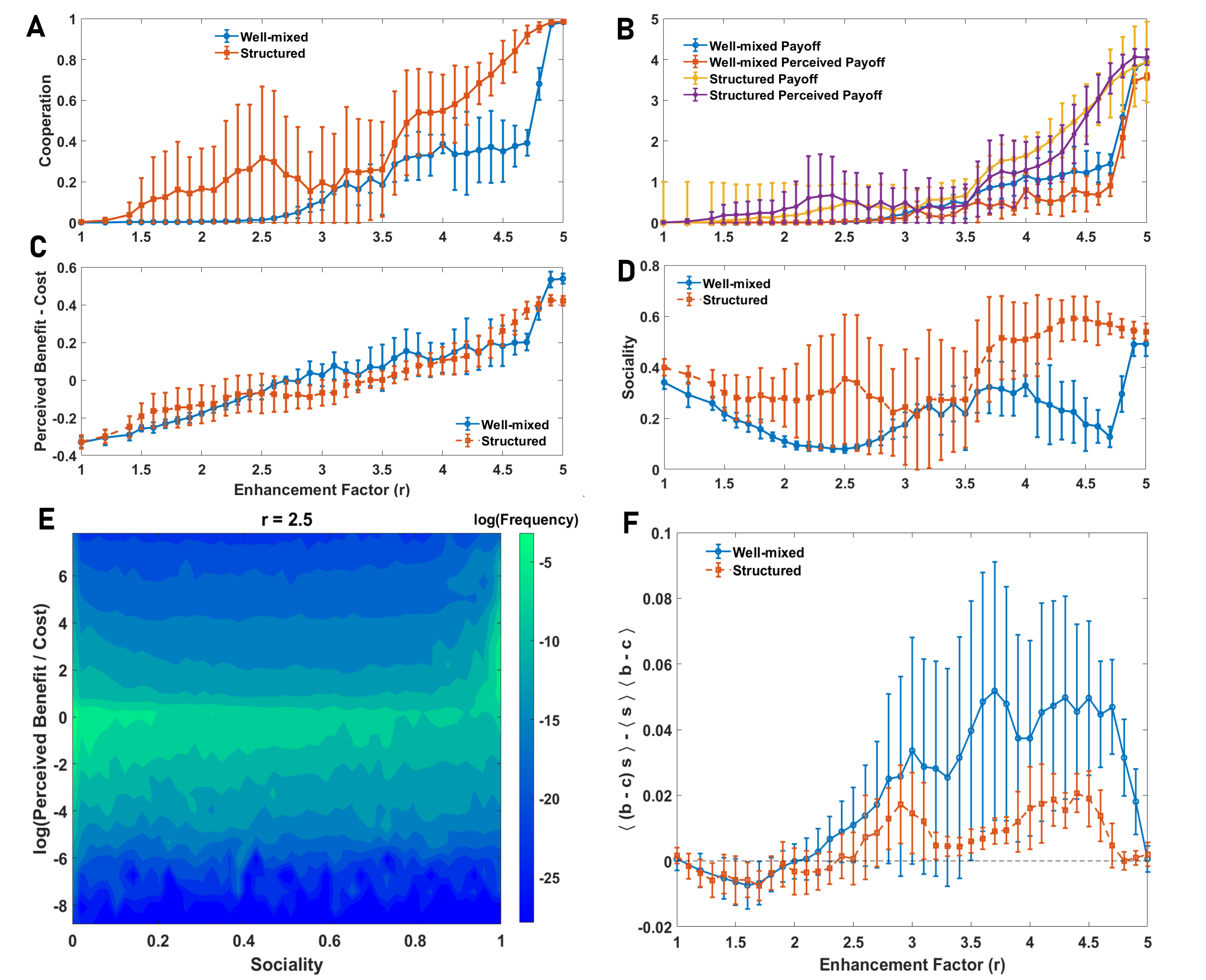}
	\caption{The evolution of cooperation, sociality, and diversity in a social perceptual public goods game. \textbf{A} and \textbf{B}: Cooperation (\textbf{A}), payoffs and perceptions of payoffs (``happiness'') (\textbf{B}) in a social perceptual public goods game in both well-mixed and structured population as a function of enhancement factor are plotted. \textbf{C} and \textbf{D}: Perceived net benefit of the public good (benefit minus cost) and average sociality in a well-mixed and structured population are plotted. While the perceptions do not show high differences, or even higher in a well-mixed population, individuals evolve to be more social in a structured population, leading to higher cooperation. \textbf{E}: The distribution of the logarithm of benefit-to-cost ratio and sociality is plotted in logarithmic scale. High social and perceptual diversity exists in the system. However, the distribution of sociality exhibits two modes for too-small and too-large sociality. \textbf{F}: The connected correlation function of sociality and perceived net benefit of the public good, $\langle (\hat{b}-\hat{c})s\rangle-\langle s\rangle \langle (\hat{b}-\hat{c})\rangle$, where, $\langle$.$\rangle$ represents an average over the population, as a function of the enhancement factor is plotted. For intermediate values of enhancement factors, where the population is behaviorally heterogeneous, a strong correlation between perceptions and sociality is observed. Parameter values: $\nu=10^{-3}$ and $N=10000$.  }
	\label{Fig3}
\end{figure}

To address this question, we employ the social perceptual public goods game, where in addition to evolving private perceptions of public good cost and benefit, individuals can evolve a social trait, $s\in[0,1]$ which determines how much they incorporate the perception of others into their own perception. $s=0$ means the individual learns privately and $s=1$, means they adopt the perception of others in their group and make collective decisions. In Fig. \ref{Fig3}\textbf{A}, we show that cooperation evolves in both well-mixed and structured populations. Higher cooperation is observed in a structured population, leading to higher payoffs and higher perception of payoff (Fig. \ref{Fig3}\textbf{B}). We note that perception of the payoff is defined as the gain from the public good based on the individuals' collective perception of public good cost and benefit. That is, for an individual, $i$, with the collective perception of public good cost and benefit, respectively, $\hat{c}^s_i$ and $\hat{b}^s_i$, and in a group with $n_C$ other cooperating individuals, the perception of the payoff is $\hat{\pi}(\sigma_i)=\delta_{\sigma_i,C}(\hat{b}^s_i-\hat{c}^s_i)+n_C\hat{b}^s_i$, where $\sigma_i\in{C,D}$ is the strategy of the focal individual.

The perception of net public good benefit (defined as the population average $\hat{b}-\hat{c}$) in a well-mixed and structured population are nearly similar (Fig. \ref{Fig3}\textbf{C}). The notable difference between a well-mixed and structured population is the evolution of higher sociality in a structured population (Fig. \ref{Fig3}\textbf{D}). While in a well-mixed population perception of the payoff is always smaller than the actual payoff, in a structured population it can be the other way. This provides two predictions: Firstly, higher sociality (regarding social information use) leads to higher cooperation. That higher sociality promotes cooperation is consistent with some past work, according to which different aspects of sociality, for instance, group living \cite{Garcia2013,Purcell2012}, are key to cooperation. Secondly, higher sociality leads to a more positive perception of payoffs (higher than the actual payoffs). A similar conclusion holds in smaller population sizes, where higher sociality evolves and leads to higher perception of payoffs (\red{S.5}). This conclusion matches empirical evidence according to which there is a strong and universal relation between diverse aspects of prosociality and happiness \cite{Chen2024,Aknin2013,Nelson2016,Dakin2022}.

In contrast to many models based on bounded rationality, full cooperation is not observed in a broad range of enhancement factors. This phenomenon, which aligns with empirical observations, for instance, in public goods experiments \cite{Chaudhuri2011}, indicates that individuals exhibit perceptual diversity, leading to behavioral diversity. To see this, in Fig. \ref{Fig3}\textbf{E} we present the joint distributions of the logarithm of perceived benefit-to-cost ratio ($\log(b/c)$) and social susceptibility for a fixed value of enhancement factor. Individuals show high perceptual and social diversity. While some evolve to perceive the public good to be highly beneficial, others perceive it to be highly costly. Furthermore, individuals show stark differences in their sociality, ranging from individuals who do not rely on social information and make private decisions to highly social individuals, who integrate the perspective of others to make decisions.

\begin{figure}[!hbt]
	\centering
	\includegraphics[width=1\linewidth, trim = 0 0 5 4, clip,]{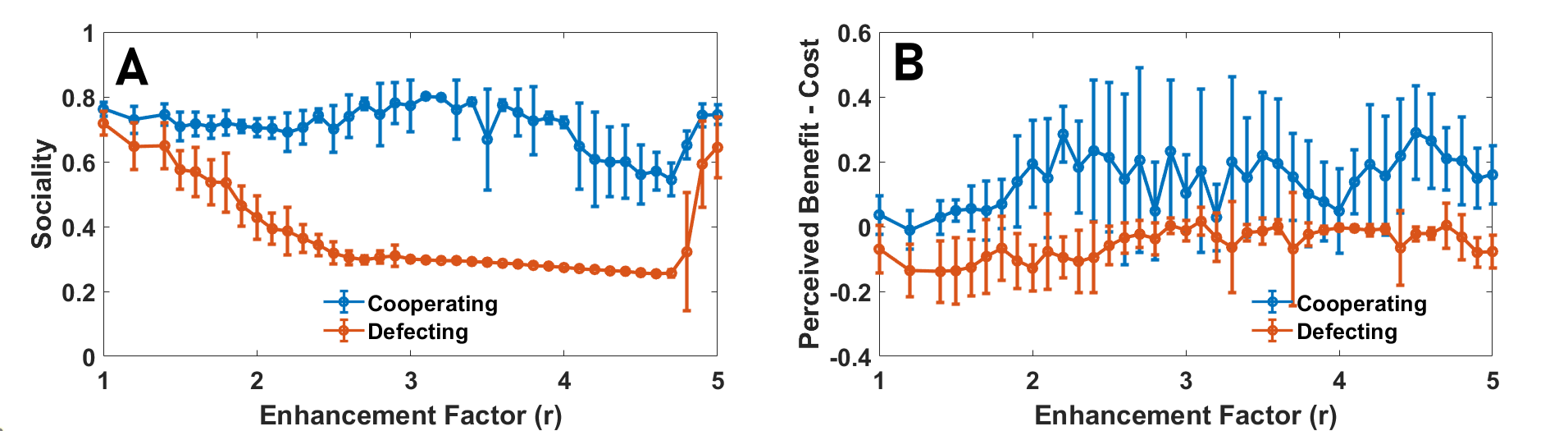}
	\caption{Social information use and perception of cooperating and defecting individuals. Sociality (\textbf{A}) and the perception of net public good benefit per individual, $\hat{b}-\hat{c}$, (\textbf{A}) separately for cooperating and defecting individuals in a well-mixed population as a function of enhancement factor is plotted. Cooperating individuals exhibit higher sociality and a more positive economic perception. Parameter values: $\nu=10^{-3}$ and $N=10000$.  }
	\label{Fig4}
\end{figure}

Notably, a strong correlation between individuals' economic perception and their sociality is observed, which indicates that individuals evolve consistent personalities over two different dimensions of behavior. Those who perceive the public good to be highly beneficial evolve higher sociality and those who perceive the public good to have higher cost than benefit evolve low sociality. To investigate how this result depends on the enhancement factor, in Fig. \ref{Fig3}\textbf{F}, we present the connected correlation function of sociality and perceived net benefit of the public good. A strong positive correlation is observed for intermediate enhancement factors, where the population is heterogeneous, and it approaches zero for too-small or too-large enhancement factors, where the population becomes behaviorally homogeneous. Besides, these personality attributes are correlated with behavior. This can be seen in Fig. \ref{Fig4}\textbf{A}, where we plot the sociality of cooperating and defecting individuals. More social individuals cooperate more. Similarly, those with a higher perception of public good benefit cooperate more (Fig. \ref{Fig4}\textbf{B}). 

This unusual diversity arises out of rational decision-making. Rather than exhibiting a rigid behavior, (perceptually) rational individuals change behavior depending on their group. In contrast to a purely rational agent, which is unique, infinitely many such perceptually rational agents exist, whose behavior varies continuously as a function of perceived net public good benefit and sociality (see Fig. \ref{Fig1}\textbf{C}). The diversity observed in the evolutionary dynamics shows that, rather than converging to a unique rational agent, evolution explores many such possibilities, giving rise to the coexistence of diverse perceptually rational personalities. This is consistent with the observation that even a simplified version of the model with binary traits does not admit an evolutionarily stable strategy. Furthermore, the very assumption of (perceptual) rationality (maximizing perceived payoffs), gives rise to the self-organized emergence of diverse \textit{personality types} who employ consistent strategies and exhibit correlations between different dimensions of behavior, which in our simple model, are limited to the social information use and economic perceptions (\red{see S.7}).

\subsection{The structure of diversity}

The bimodality of the distribution of sociality, apparent in Fig. \ref{Fig3}\textbf{E}, invites a naive classification of personalities into two classes, collectivistic individuals who excessively rely on the collective perception of others and often evolve a pro-social economic perception, and individualistic perspectives, who evolve low sociality and often a less pro-social or anti-social economic perception. However, many individuals do not belong to any of the two extremes, as evolved personalities, are in fact, a continuum of personalities from highly defecting to highly cooperating individuals, suggesting despite its apparent simplicity, a genuine diversity resulting from complex evolutionary dynamics of rational agents is at work in our ecosystem. 

\begin{figure}[!hbt]
	\centering
	\includegraphics[width=1\linewidth, trim = 30 0 25 0, clip,]{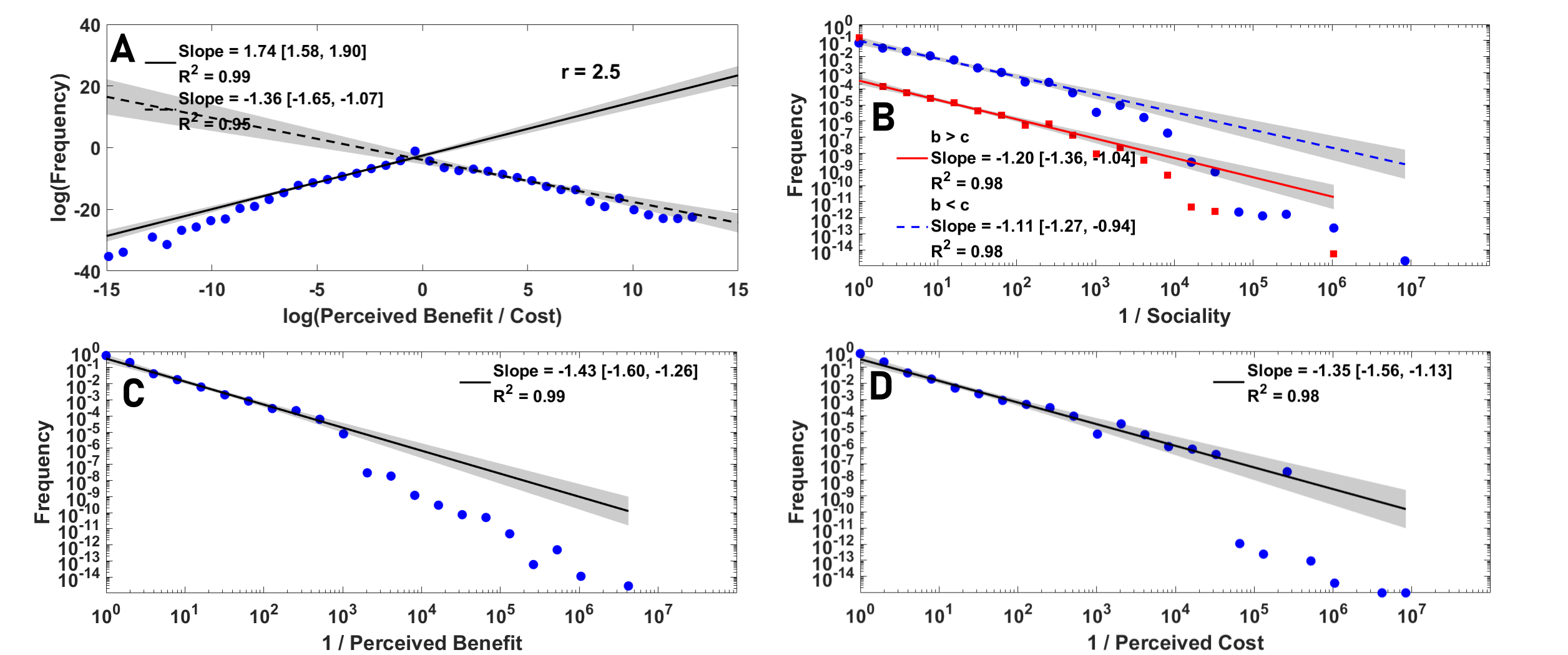}
	\caption{The structure and function of diversity. \textbf{A}: The logarithm of the frequency of perceived benefit-to-cost ratio as a function of the logarithm of benefit-to-cost ratio is plotted. The perceived benefit-to-cost ratio exhibits an asymmetric power law with an exponent between $-1$ and $-2$, depending on the enhancement factor. \textbf{B}: Inverse sociality exhibits a power law tail, with separate laws for personal ($b>c$) and anti-social ($c<b$) perceptions, with exponents close to $-1$. \textbf{C} and \textbf{D}: The inverse perceived benefit and perceived cost exhibit a power law tail with an exponent between $-1$ and $-2$. Social information use and perception of cooperating and defecting individuals. Sociality (\textbf{A}) and the perception of net public good benefit per individual, $\hat{b}-\hat{c}$, (\textbf{A}) separately for cooperating and defecting individuals in a well-mixed population as a function of enhancement factor is plotted. Cooperating individuals exhibit higher sociality and a more positive economic perception. Parameter values: $\nu=10^{-3}$ and $N=10000$.}
	\label{Fig5}
\end{figure}

We begin studying the structure of diversity, by discussing the \textit{behavioral power-laws} governing perceptual and social diversity. The exponents of some behavioral power laws exhibited by the model are summarized in Table \ref{Table1} (\red{see S.4 for details}). The perceived benefit-to-cost ratio, and the inverse sociality separated for those who perceive the public good to be (individually) beneficial, $b>c$ (which can be called, pro-social economic perceptions), and those who perceive it to be individually detrimental, $b<c$ (anti-social perceptions), in a well-mixed population are presented in Figs. \ref{Fig5}\textbf{A} and \ref{Fig5}\textbf{B}, respectively. A similar tent-shaped power law to that observed in a perceptual public goods game is observed. However, the exponents are not universal and exhibit meaningful variations with parameter values, which contain important information about the dynamics. The power-laws approach a symmetric power-law with an exponent close to $2$, for too-large and too-small enhancement factors, where the dynamics exhibit behavioral homogeneity and converge to either full defection or full cooperation. In contrast, it becomes asymmetric and shows smaller exponents in between, indicating higher diversity. Similar phenomena are observed, for inverse perceived cost and benefit, presented in Fig. \ref{Fig5}\textbf{C} and \ref{Fig5}\textbf{D}.

Perceptual diversity drives social diversity. Furthermore, the fact that social diversity is driven by the perceptual diversity of the individuals leads to differentiated ``laws'' governing the sociality of individuals with dramatically different perceptions. Inverse sociality of both pro-social and anti-social perceptions exhibit a power law with a lower exponent for anti-social perceptions ($b<c$), indicating a faster decline of sociality in these individuals compared to pro-social perceptions. This is consistent with the correlation we observed between economic perceptions and sociality. While the exponent is close to $-2$ among pro-social perceptions, implying a nearly uniform distribution for sociality it is closer to $-1$ for anti-social perceptions, implying an exponent close to $-1$ for sociality. This value corresponds to a Zipf law containing maximum diversity and complexity \cite{Marsili2022}. This results from the fact that these individuals use diverse forms of conditional strategies.

These results confirm our expectation that the perceptual diversity observed in a simple perceptual public goods game can only increase when the door is open to the evolution of social structure, leading to enhanced perceptual diversity, which in turn triggers social and behavioral diversity.

\subsection{Functional role of diversity: A simplified model with binary traits:}

So far we have mainly focused on the structure of diversity. An essential question is the functional role of diversity. This is intimately related to the evolution of consistent rational personalities. Different mechanisms are suggested for the evolution of consistent personalities \cite{Dingemanse2010,Wolf2012,Wolf2008,Wolf2008a,Biro2008,Rands2003,Stamps2007,Dall2004,Salahshour2024,McNamara2009,Bergmüller2010,Salahshour2021,Botero2010} (see the \nameref{Discussion}). However, our work demonstrates that an important mechanism that has so far been overlooked is rational decision-making. This is the case because rational decision-making is by nature flexible and allows individuals to change their behavior based on their environment in a way that is consistently determined by individuals' perceptions and sociality, and thus, exhibits contextual consistency, which is essential to a strict notion of personality \cite{Kaiser2021}. While a power law distribution of perceptions and sociality indicates the diversity of strategies is genuine and a classification of the individual into few categories is rather arbitrary, to shed more light on the functional role of diversity, and diverse rational personalities, we appeal to an even simpler model, where individuals can only evolve binary traits: a binary perception of the public good cost, benefit, and a binary social trait. As we showed before, no type is an evolutionarily stable strategy in this simplified model, leading to an ever-shifting polymorphism in the evolutionary dynamics. We use both simulations and replicator dynamics to study this simplified model (\red{see \nameref{Methods} and S.8}).

\begin{figure}%[ht]
	\centering
	\includegraphics[width=1\linewidth, trim = 110 304 90 10, clip,]{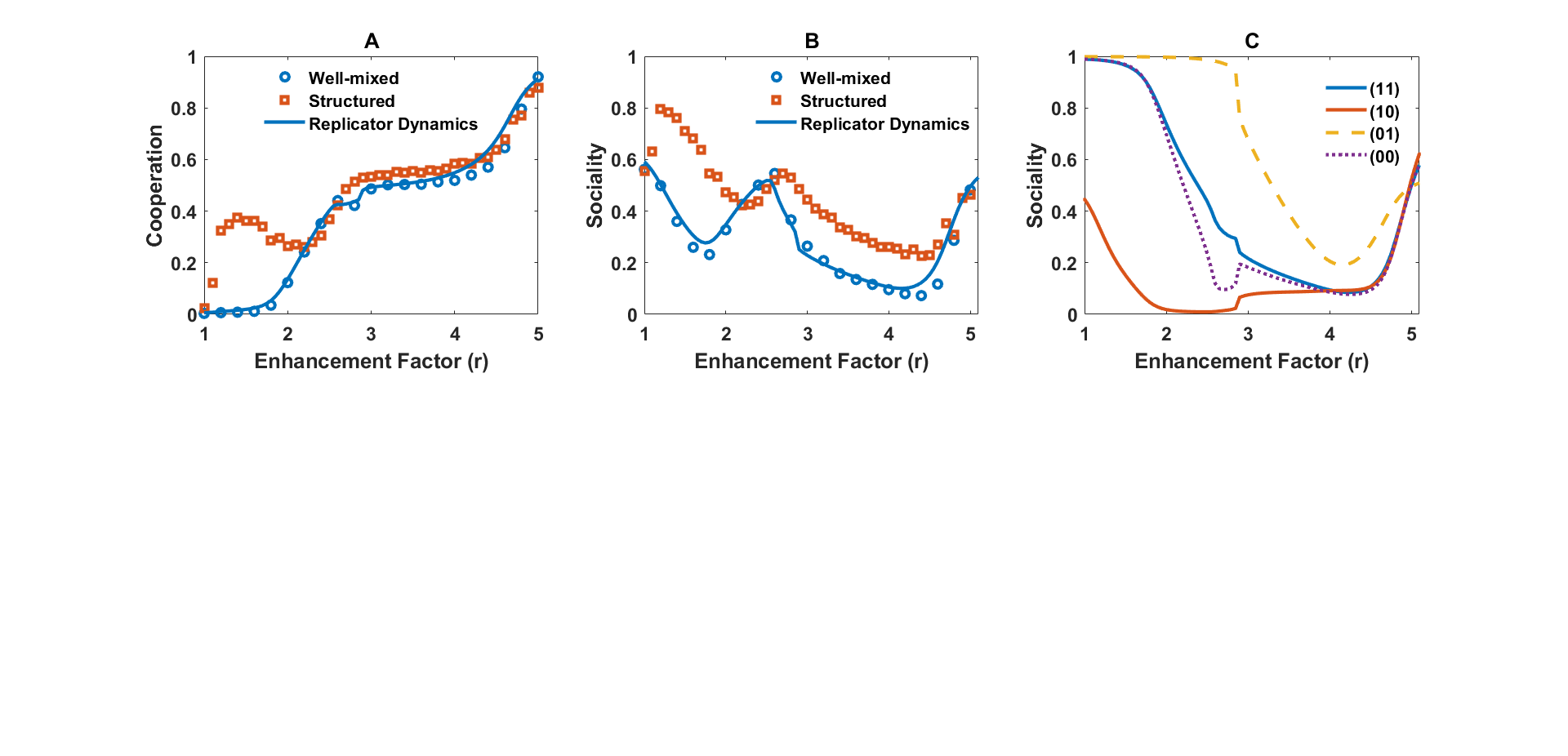}
	\caption{The evolutionary dynamics of the model with binary traits. Cooperation (\textbf{A}) and sociality (\textbf{B}) in a simplified version of the model with the binary perception of cost and benefit and sociality in both well-mixed and structured populations are plotted. Reduced diversity leads to higher cooperation, indicating diversity is detrimental to cooperation in rational individuals. \textbf{C}: The fraction of social individuals among four economic perceptions of public good cost and benefit as a function of enhancement factor is plotted. Pro-social economic perceptions exhibit higher sociality, indicating a consistency of two dimensions of personality. Furthermore, individuals with diverse personalities coexist in the population and specialize in different behavioral strategies. Parameter values: $\nu=10^{-3}$, in a well-mixed population, $N=25600$ and in a structured population $N=6400$. }
	\label{Fig6}
\end{figure}

Cooperation in both well-mixed and structured populations, as well as replicator dynamics results (corresponding to a well-mixed population in the limit of infinite population size), is plotted in Fig. \ref{Fig6}\textbf{A}. In a broad range of enhancement factors, higher cooperation than that observed in the model with continuous perceptions and sociality is observed. This is the case for small enhancement factors where the evolution of cooperation is a strong social dilemma. This indicates that diversity can be detrimental to cooperation when social dilemmas are too strong. This result is consistent with many empirical observations \cite{Baldwin2010,Habyarimana2007,Hjort2014,Dinesen2020,Banerjee2005}, however, contrasts most of the work in evolutionary game theory \cite{Santos2008,Shi2010,Zhang2010,Liu2017,Zhu2014,Santos2012,Su2016,Van2009,Perc2008,Pacheco2009,Szolnoki2007,Gao2010} (see the \nameref{Discussion}). Furthermore, our work suggests that diversity can play a constructive role for public good cooperation, when the social dilemma is weak (large enhancement factors), especially in a structured population. The question that how the effect of diversity on public good provision depends on the strength of social dilemma does not seem to have been investigated empirically.

\begin{figure}%[ht]
	\centering
	\includegraphics[width=1\linewidth, trim = 90 4 90 10, clip,]{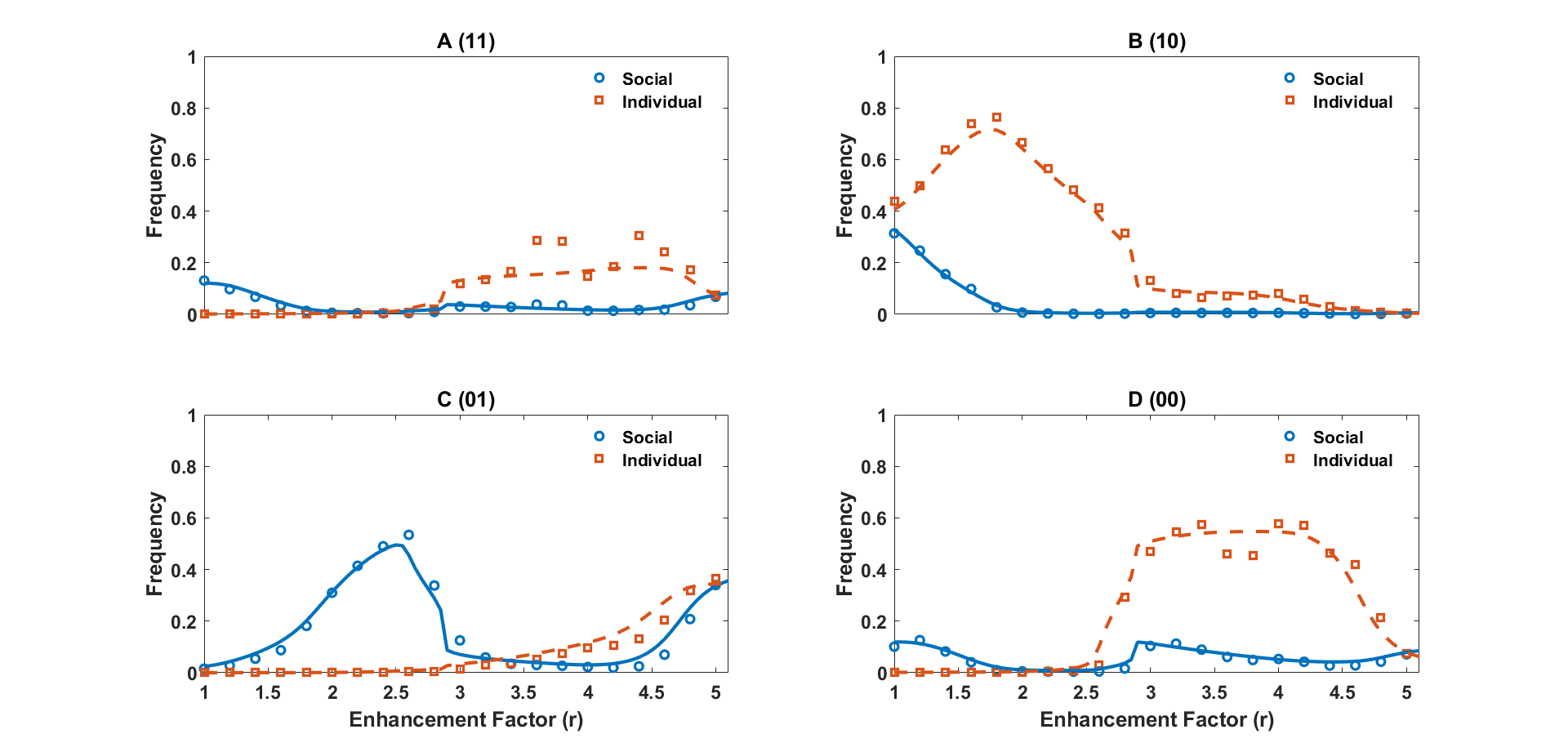}
	\caption{ The frequency of different economic perceptions and sociality in a simplified version of the model with binary perceptions of cost and benefit and sociality are plotted. Each panel shows social and asocial individuals for one of the economic perceptions. Lines show replicator dynamics results and markers show simulations in a well-mixed population. Parameter values: $\nu=10^{-3}$. In a well-mixed population, $N=25600$.} 
	\label{Fig7}
\end{figure}

In this simplified model, individuals can evolve only four perceptions, which we call Neutral (zero cost and benefit for the public good, $00$), Public good (unit cost and benefit, $11$), Harmony (zero cost and unit benefit, $01$), and Anti-harmony (unit cost and zero benefits, $10$) perceptions. Besides, each type can be social or asocial. In Fig. \ref{Fig6}\textbf{B}, we present the fraction of social individuals in the population in both well-mixed and structured populations, together with the replicator dynamics. In \ref{Fig6}\textbf{C}, we present the fraction of social individuals in each type. This is the sociality of different economic perceptions. Fig. \ref{Fig7} presents the frequency of each type in the population as a function of the enhancement factor (both finite well-mixed population and replicator dynamics).

Consistently with the lack of evolutionarily stable strategy, in the evolutionary dynamics, no type can successfully exclude other types. This is because individuals evolve to specialize in different economic perceptions and social information use strategies, leading to different adaptive behavioral types. Furthermore, individuals evolve consistent personalities over two dimensions: those with prosocial economic perceptions are more likely to evolve to rely on social information, form collective personalities, and make collective decisions, and those with anti-social economic perceptions are more likely to be individualistic, form private beliefs and make decisions individualistically (Fig. \ref{Fig6}\textbf{C}). This indicates a consistency of different dimensions of rational individuals' personality which evolves naturally and in a self-organized way. Furthermore, we observe individuals who employ a fixed strategy (asocial Harmony and Anti-harmony), those with responsive behavior who switch strategy based on context (social individuals), or those who randomize their strategy (asocial neutral and public good perceptions), exhibiting a form of bet-hedging. This indicates rational individuals evolve to exploit all the possible strategies. This role specialization results from the flexibility of behavior: the fact that rational individuals change their behavior based on their social context. As mentioned before, this is also essential to the lack of an evolutionary stable strategy, as each type, while may perform well in the presence of some other types, underperforms in competition with others. Namely, the existence of randomizing strategies, who exploit antisocial perceptions but perform worse than prosocial perceptions is the key to the lack of evolutionary stable strategy and maintaining behavioral diversity in the population. In the Supplementary Information, we investigate how the complexity of rational decision-making leads to such a rich phenomenology (\red{see S.8}).

\subsection{The effect of group size, non-monotonic evolution, and cyclic dominance}

So far, we have focused on group size equal to $5$. This choice is motivated by the observation that most of the previous studies in the evolutionary game theory of public goods games have adopted this choice, and it is generally accepted that the evolutionary dynamics exhibit reasonable robustness with respect to the variation of the group size \cite{Perc2013}. However, the deviation of our framework from the usual bounded rationality approach may raise the question that to what extent group size affects the results. In this section, we address this question.

\begin{figure}
	\centering
	\includegraphics[width=1\linewidth, trim = 110 20 110 20, clip,]{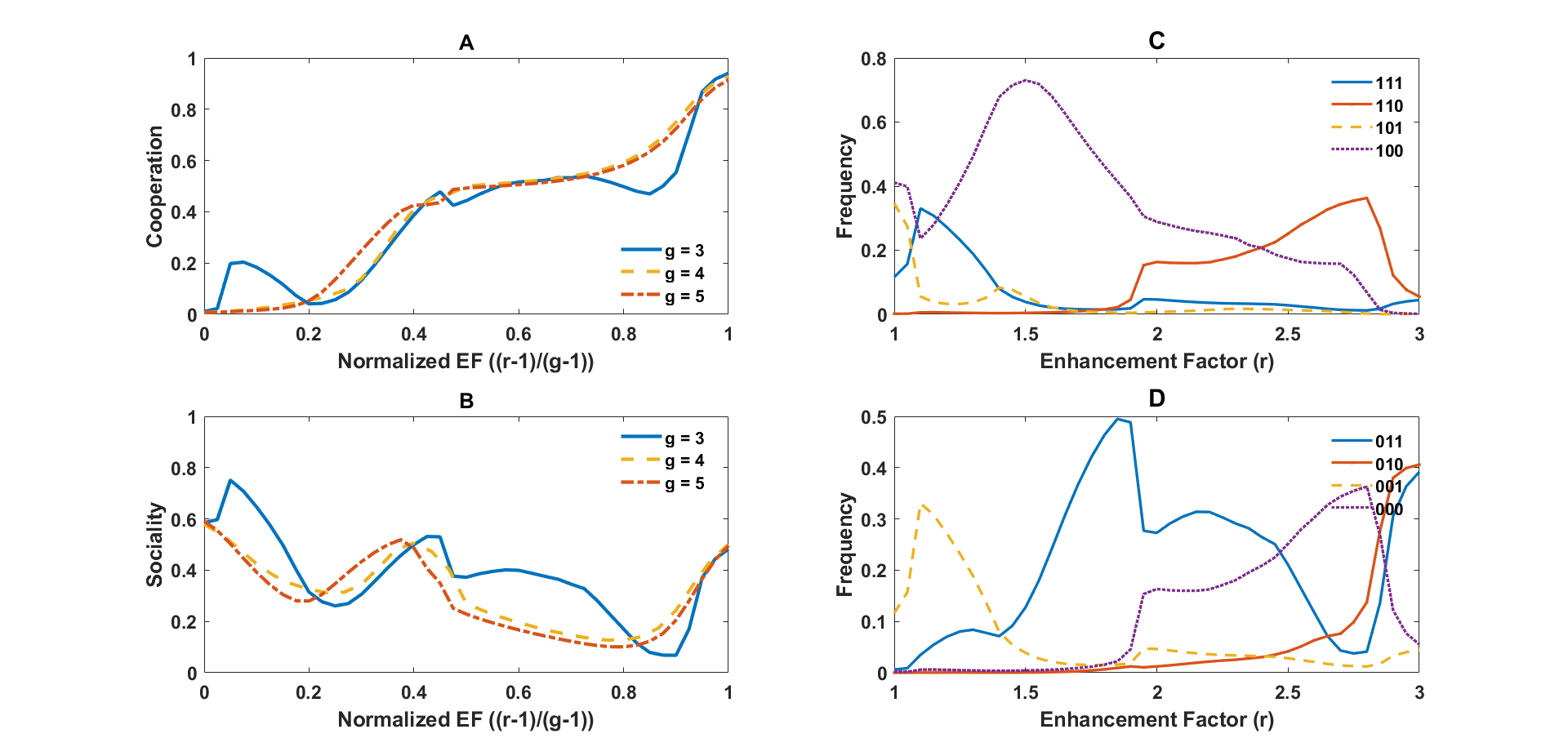}
	\caption{The effect of group size in a binary social perceptual public goods game. \textbf{A} and \textbf{B}: The time average cooperation (\textbf{A}) and sociality (\textbf{B}) as a function of the normalized enhancement factor in a binary social perceptual public goods game and in groups of various sizes are plotted. Here, the replicator dynamic is used. Up to some shifts in the values of enhancement factors for which cooperation evolves, the results in larger groups are largely similar. However, for $g=3$, cooperation can evolve even for enhancement factors close to $r=1$. Besides, while in larger groups, the dynamics settle in a fixed point, for $g=3$ cyclic dominance of different strategies leading to periodic orbits is observed in a range of enhancement factors (normalized enhancement factor, approximately between $0.5$ and $0.9$, corresponding to two local minima in cooperation level for $g=3$). \textbf{C} and \textbf{D}: The time average frequency of different strategies for $g=3$ as a function of the enhancment factor are plotted. Here, replicator dynamics results are used. Parameter values: $\nu=10^{-3}$. }
	\label{Fig8}
\end{figure}

In Figs. \ref{Fig8}\textbf{A} and \ref{Fig8}\textbf{B}, we plot cooperation and sociality as a function of the normalized enhancement factor in a binary social perceptual public goods game in groups of $g=3$, $g=4$, and $g=5$ individuals, using the replicator dynamics. Here, we use a normalized enhancement factor to keep the results for different group sizes comparable. We note that we keep the group size beyond $2$ so that the chosen group size satisfies the assumption of a public goods game (which is supposed to take place in groups larger than two individuals \cite{Perc2013,Szabo2007}). Up to small shifts in the values of enhancement factor for which cooperation evolves, the results are largely similar for larger group sizes. A more dramatic change is observed for $g=3$. In this case, cooperation can evolve even for enhancement factors slightly larger than $1$. This is due to the fact that rational individuals can reciprocate against defective strategies more easily and without withholding cooperation from more optimistic perceptions in smaller group sizes. Besides, both cooperation and sociality show a non-monotonic behavior as a function of the enhancement factor. This phenomenon may be surprising given that such a nonmonotonic behavior has not been typically observed in diverse modifications of public goods games in the bounded rationality framework. However, this phenomenon, also present in other variants of the model (although often to a smaller degree), is a feature of our perceptual rationality framework, not a bug. This results from the fact that different rationality types change their behavior based on local information and group composition (in other words, the flexibility of rational decision-making).

To see this, in Figs. \ref{Fig8}\textbf{C} and \ref{Fig8}\textbf{D}, we plot the time average frequency of different types as a function of the enhancement factor in groups of $g=3$ individuals. A look at the frequency of different strategies reveals that cooperation observed for small enhancement factors is due to an increase in the frequency of social randomizing types, $(001)$ and $(111)$ at the sake of anti-harmony types, $(100)$ and $(101)$. As mentioned before, these strategies perform better than highly defective $(10.)$ types, due to cooperating in groups with a positive perception of public good and defecting in groups with a negative perception of public good. An increase in the frequency of social randomizing types gives an advantage to social harmony perception, $(011)$, whose frequency starts to increase due to selectively cooperating in groups composing randomizing types. This, in turn, leads to an increase in the frequency of the individualistic anti-harmony type, $(100)$, the frequency of which increases in this region (due to exploiting the harmony type) and slows down the growth of the social harmony type. This leads to a decrease in overall cooperation up to the enhancement factor, $r\approx1.5$. Notably, in this region, the social anti-harmony type does not perform well compared to the individualistic anti-harmony type, indicating that social information use is not an adaptive strategy for such individuals with an anti-social perception of the public good (leading to the evolution of contextually consistent personalities).

For larger enhancement factors, the social harmony type increases in frequency due to obtaining a higher payoff in small groups of like-minded individuals. This leads to an increase in cooperation and sociality in this region. At an enhancement factor close to $2$, individualistic randomizing types, $(110)$ and $(000)$ start to obtain a benefit and increase in frequency. This is associated with a drop in the frequency of social harmony type. An increase in the frequency of individualistic randomizing types in this region results from the fact that because of their neutral perception, they do not affect the collective perception of the group. Thus, social harmony types do not withhold cooperation from these types, while they do withhold cooperation in groups where anti-harmony perceptions have the majority. Due to the existence of randomizing types, cooperation in the system remains almost constant up to an enhancement factor close to, approximately, $2.7$, while sociality decreases in this region due to the increasing frequency of randomizing types. This decreased sociality offsets the increase in the enhancement factor, leading to nearly constant cooperation as the enhancement factor increases. For larger values of enhancement factors, cooperation is beneficial enough and both social and individualistic harmony perceptions increase in frequency, leading to an increase in both cooperation and sociality.

This discussion shows that the \textit{non-monotonic evolution} typically observed in the evolutionary dynamics of perceptually rational agents results from their complex dynamics. This complex dynamics, in turn, originates from the lack of evolutionarily stable strategies, leading to the evolution of diverse rationality types, that engage in complementary and/or competitive roles. Consequently, improvements in external economic productivity (i.e., enhancement factor), do not necessarily lead to an increase in cooperation (and sociality). Rather, counter-intuitively it can even lead to a decrease in cooperation (or sociality), due to radical shifts in the population configuration leading to the emergence of different sets of competitive, mutualistic, or complementary relations in the system.

\begin{figure}
	\centering
	\includegraphics[width=1\linewidth, trim = 110 20 110 20, clip,]{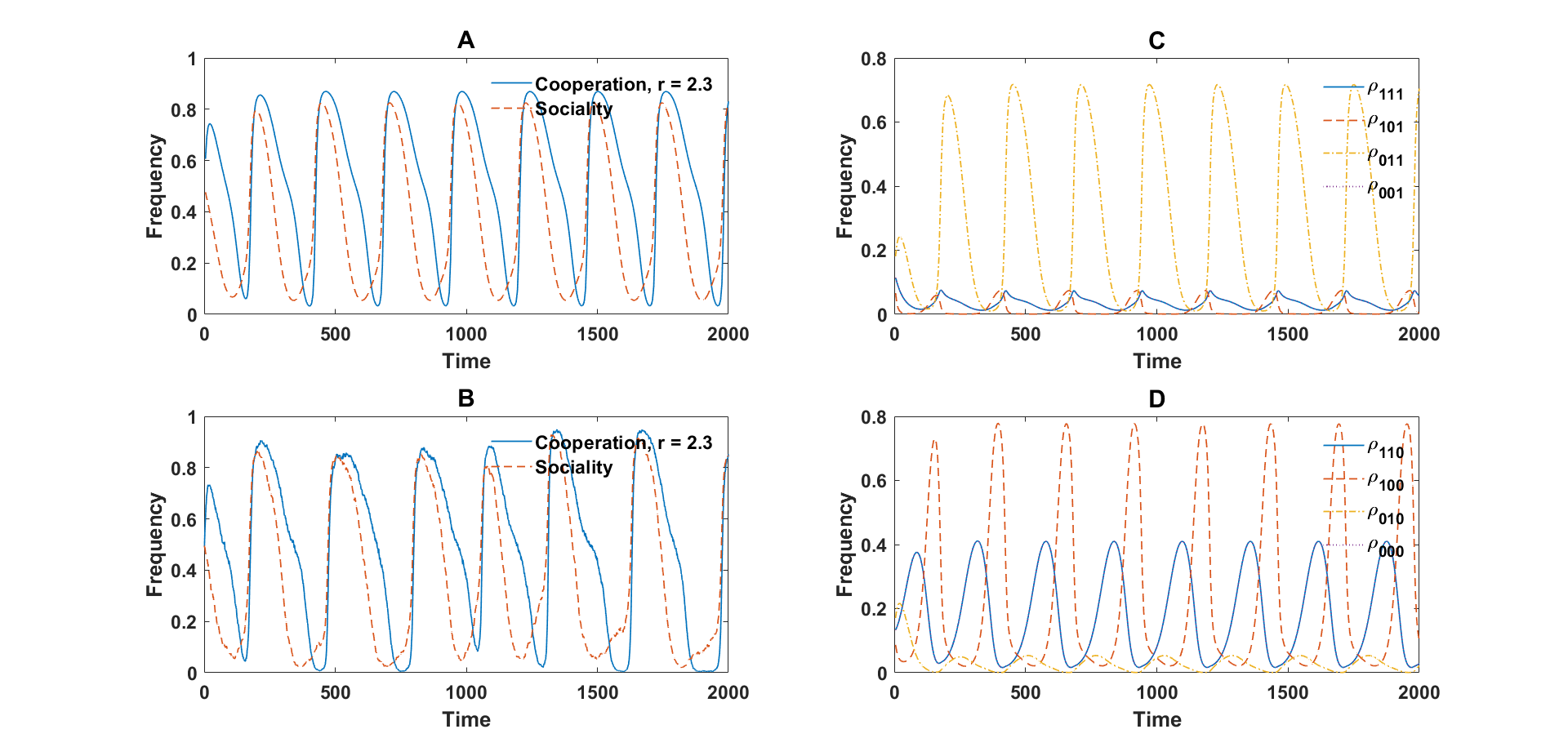}
	\caption{Cyclic dominance of different types in a binary social perceptual public goods game in groups of $g=3$ individuals. \textbf{A} and \textbf{B}: Cooperation and sociality for $r=2.3$ and $g=3$ resulting from the replicator dynamics (\textbf{A}) and a simulation in a population of $N=10^4$ individuals (\textbf{B}) as a function of time are shown. The dynamics go through cycles of high and low sociality, followed by high and low cooperation, resulting from the cyclic dominance of different types. \textbf{C} and \textbf{D}: The frequency of different types as a function of time are shown. Here, replicator dynamics results are used. Parameter values: $\nu=10^{-3}$ and $g=3$.}
	\label{Fig9}
\end{figure}

Notably, in small group sizes ($g=3$), the evolutionary dynamics can exhibit periodic orbit, resulting from the cyclic dominance of different strategies. Such periodic orbits are observed in the region where individualistic randomizing types are found in the population with high frequency (between two local minima in the cooperation level in Fig. \ref{Fig8}\textbf{A}). Examples of such periodic orbits are presented in Fig. \ref{Fig9}, where similar dynamics in replicator dynamics and simulations in finite populations are observed. Such periodic orbits are not observed in groups of $g=4$ individuals or higher. The emergence of periodic orbits in the evolutionary dynamics in smaller group sizes can be understood by noting that the probability of the formation of homogeneous groups increases for smaller group sizes. This can result in smaller payoff differences between individuals with the same type (who less frequently happen to play their game in groups with different compositions, when group size is smaller), leading to larger average payoff differences between individuals with different types, which can drive the cyclic dominance of different strategies. By contrast, in large groups, heterogenous groups are formed more frequently. Consequently, individuals with the same type obtain different payoffs in different groups. This leads to the coexistence of different types rather than their cyclic dominance over time.

\subsection{Evolvable perception of only cost or only benefit of the public good}

To shed more light on the evolutionary dynamics, we consider two other simplifications of the social perceptual public good game evolutionary dynamics, where, while the traits are still continuous, individuals have perfect information of either public good cost or benefit, but their perception of the public good benefit or cost, respectively, is evolvable. To accommodate the possibility that individuals can potentially perceive the public good to be beneficial or detrimental, we limit the evolvable trait in the interval $[0,2]$.

\begin{figure}
	\centering
	\includegraphics[width=1\linewidth, trim = 130 0 105 24, clip,]{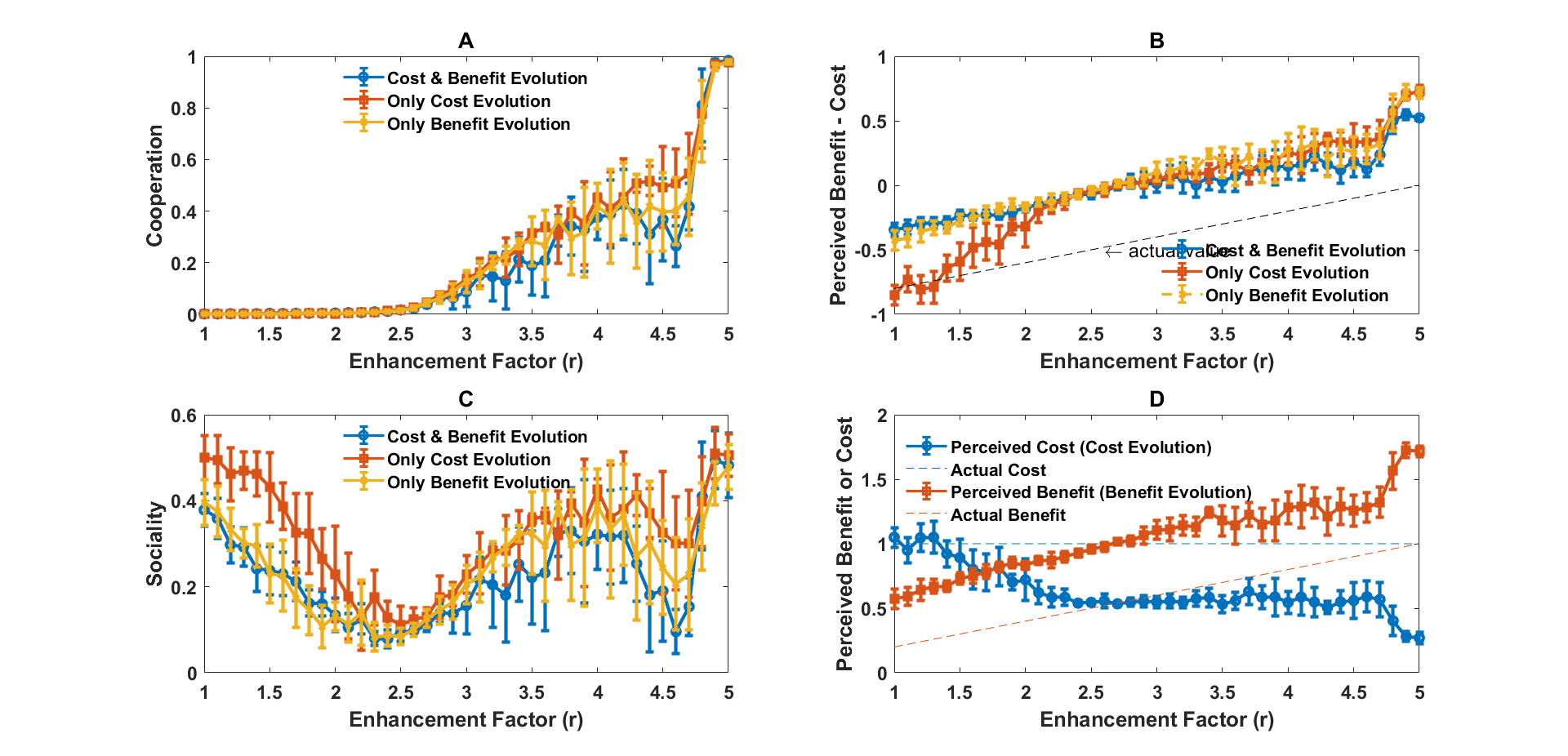}
	\caption{The evolutionary dynamics of the social perceptual public goods game with only evolvable cost or benefit perception. Cooperation (\textbf{A}), population average of public goods net benefit, $\hat{b}-\hat{c}$, (\textbf{B}), sociality (\textbf{C}), and public goods cost or benefit \textbf{D}, are shown in models where individuals have access to the objective values of public good cost or benefit and only their perception of, respectively, public good benefit or cost is subject to evolution. For comparison, in \textbf{A} to \textbf{C}, we have also shown cooperation, net perceived public good benefit ($\hat{b}-\hat{c}$), and sociality when both public good cost and benefit are evolvable. Higher cooperation and higher sociality are observed in simpler evolutionary dynamics with only one evolvable perceptual trait. Parameter values: $\nu=10^{-3}$ and $N=6400$. }
	\label{Fig10}
\end{figure}

In Fig. \ref{Fig10}\textbf{A}, we compare cooperation in a well-mixed population in the three evolutionary dynamics. That is, when only the perception of public good cost is evolvable, when only the perception of public good benefit is evolvable, and when the perception of both public good cost and benefit are evolvable. We observe higher cooperation in the simplified evolutionary dynamics with only one evolvable perceptual trait. The higher cooperation is associated with higher perception of public good net benefit (Fig. \ref{Fig10}\textbf{B}) and higher sociality of individuals (Fig. \ref{Fig10}\textbf{C}). Consequently, in simplified evolutionary dynamics, individuals evolve more optimistic perceptions (which deviate more from reality) leading to higher sociality, cooperation, and payoffs. This is consistent with our results that simpler evolutionary dynamics favor cooperation due to de-escalating the evolutionary race between different rationalities. Besides, we observe that for enhancement factors approximately between $3.5$ and $4.7$, cooperation when only public good cost is evolvable is higher than that when only public goof benefit is evolvable.

In Fig. \ref{Fig10}\textbf{D}, we plot the population average perception of public good cost (when only the perception of the public good cost is evolvable) and public good benefit (when only the perception of the public good benefit is evolvable). In reality, the public good cost is always $1$, and the personal benefit of the public good is $1/g\leq r/g\leq1$ (shown in the figures). We find that the perception of public good cost and benefit typically follows the actual value, however, with an optimism bias. For public good benefit, this trend is close to linear with respect to the actual enhancement factor with a slope comparable to the actual value but with an optimistic bias (intercept) towards larger values than reality. However, in the regions between approximately $r=3.5$ and $r=4.7$, we observe a nearly constant perception of public good benefit, irrespective of the fact that the actual value of public good benefit is increasing. As we have seen, such a constant or non-monotonic trend, also observed in other variants of the model, results from the complex dynamics and shifting complementation or exploitation relationships between different rational types.

For public good cost, we observe three regimes. For too small an enhancement factor, where most of the individuals defect, the perception of public good cost is close to the actual value ($1$). However, it drops as the enhancement factor increases and cooperation starts to evolve. A constant perception of the public good cost, but below the actual cost, is observed in the intermediate regime, where cooperation and defection co-occur in the population. Because individuals have perfect information about the actual value of public good benefit, this constant perception of public good cost can explain higher cooperation observed in the region between $r=3.5$ and $r=4.7$ when only public good cost is evolvable. Finally, the perception of public good cost further drops for too high enhancement factors, where most of the individuals cooperate. See \red{S.9} for further details.

\subsection{Large-scale eco-evolutionary patterns and their modification by the evolution of social structure}

One of the aims of this paper is to establish a connection between evolutionary game theory and large-scale eco-evolutionary patterns, a relationship that has not been demonstrated in the literature so far. While it may be feasible to explore such connections within the conventional bounded rationality framework, this linkage has yet to be illustrated. In contrast, we show here that within the perceptual rationality framework, this connection readily emerges.

The evolutionary dynamics of the perceptual public goods game align with several observed large-scale eco-evolutionary patterns, providing predictions about how social structure evolution may influence these patterns. These patterns, along with their estimated exponents, consistent with empirical records, are summarized in Table \ref{Table1} (see S.10 for further details). 

The number of ``species'' originating from a common ancestor displays a power spectral density close to an exponent of $-1$, consistent with observed evolutionary data on species diversity over time \cite{Drossel2001,Sole1996,Sole1997}. Similarly, species lifespan follows a power law with an exponent near $-2$, matching empirical observations \cite{Drossel2001,Sole1997,Sole1998,Drossel1998,Pigolotti2005}. The total number of individuals over a species lifespan and the mean population size of species also exhibit power laws with exponents close to $-2.5$ and $-2.6$, respectively. Although these patterns have not been empirically examined in this exact form, species abundance distributions have been studied, showing varied distributions across small-scale ecosystems, including a power laws distribution \cite{McGill2007,Baldridge,Matthews2015}.

Finally, we examine Taylor's power law, which describes species' temporal variation in population size over their lifespan, showing an exponent close to $1$, consistent with observed empirical patterns where this exponent often lies between $0.5$ and $1$ \cite{Taylor1961,Taylor1984,Eisler2008}.

The emergence of such large-scale eco-evolutionary patterns in a simple perceptual public goods game may be surprising. However, this results from perceptual diversity, exhibited by our perceptual rationality framework, which, combined with the stochastic multiplicative processes in evolutionary dynamics, effectively captures these large-scale eco-evolutionary patterns. As mentioned before, although a connection between evolutionary game theory and large-scale eco-evolutionary patterns has not been established before, we do not aim to argue that this is not possible within a bounded rationality framework. 

The emergence of large-scale eco-evolutionary patterns also allows us to address an interesting question: how does the evolution of social structure modify such large-scale eco-evolutionary patterns? For instance, one can think of the complex collective behavior as an emergent level of biological organization, emerging out of the evolution of new ``laws'' of interaction (e.g., patterns governing the adaptive value of different behavioral types in our model, what we referred to as \textit{shifting competitive or complementary relationshipd} before), by ``breaking'' the underlying, simpler dynamics (or the symmetries defining such dynamics) observed in the absence of such social structure \cite{Levin1992}. Arguably, this process can have similarities with the way that collective behavior emerges out of symmetry-breaking phase transitions and gives rise to emergent laws in physical systems \cite{Anderson1972}.

To address how emergent social structure may modify observed eco-evolutionary regularities, we examine how the large-scale eco-evolutionary patterns exhibited by a simple perceptual public goods game are modified by a social perceptual public goods game, due to evolution of social information use. The results of our analysis are summarized in Table. \ref{Table1} (\red{see S.10}). While in a simple perceptual public goods game large-scale eco-evolutionary patterns do not depend on the parameter values and show only a minor dependence on population structure, a more dramatic dependence on parameter values is observed in a social perceptual public goods game where a social structure evolve. This parameter-dependence can result from the shifting social configuration and relationships in the population, as external conditions (parameter values) change. Thus, rather than observing \textit{universal} patterns, as we do observe in a simple perceptual public goods game, we observe ad hoc, parameter-dependent, system-specific patterns.

In many cases, the very nature of the power-laws is preserved even when a social structure evolves. However, the exponents change, reflecting important information about the dynamics. For both too-small and too-large enhancement factors, where the population becomes behavioral homogeneous, with either full defection or full cooperation, the exponents approach those observed in a simple perceptual public goods game. In contrast, for medium enhancement factors deviation in the exponents is observed due to the introduction of social structure. The effect of the evolution of social structure on large-scale eco-evolutionary patterns varies for different patterns. Only a minor change in species mean and total population size exponent over their lifespan upon introducing social structure is observed, suggesting these statistics are robust, contain universal information, and are less sensitive to details. However, a more dramatic change in species population size per time step is observed, reflecting the sensitivity of these statistics to details. 

\begin{figure}
	\centering
	\includegraphics[width=1\linewidth, trim = 30 0 35 4, clip,]{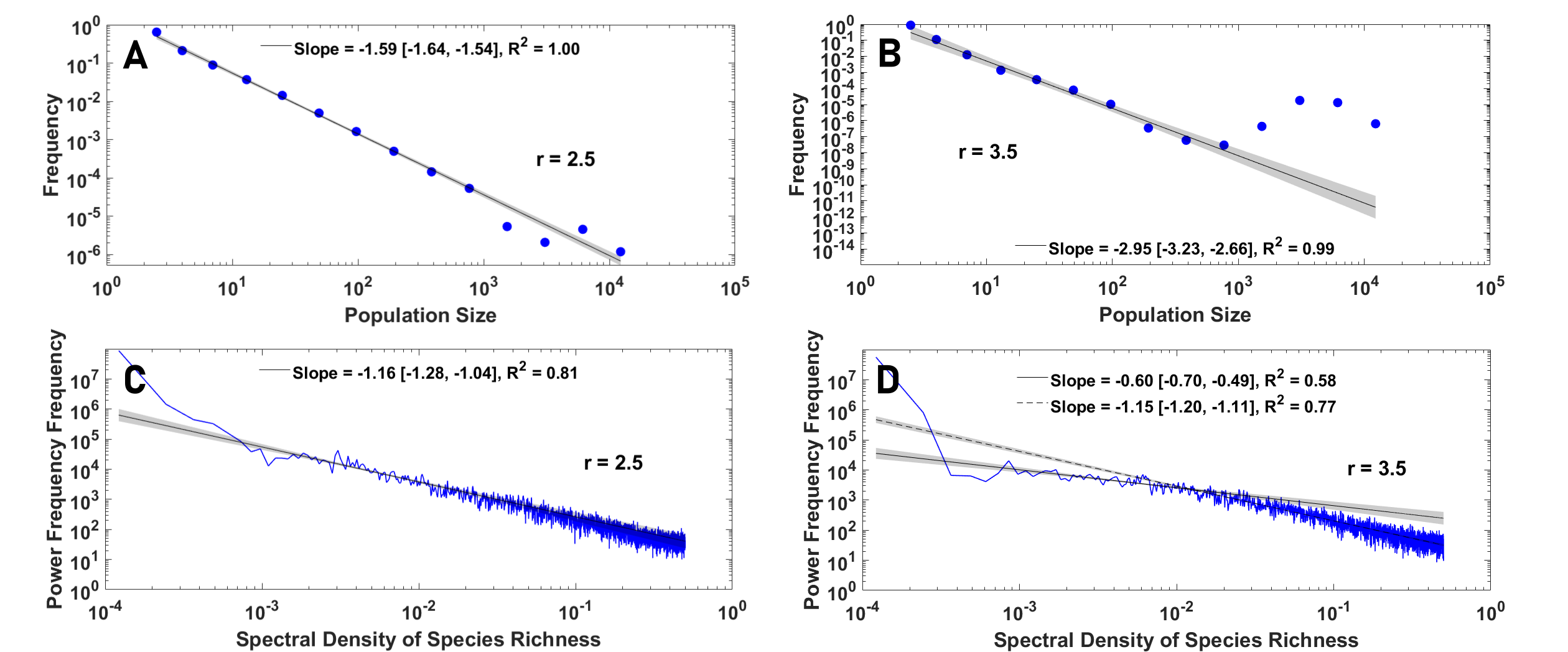}
	\caption{Modification of large-scale eco-evolutionary patterns due to the evolution of social structure. \textbf{A} and \textbf{B}: The distribution of species population size per timestep for two different values of enhancement factor ($r$) are plotted. The evolution of social structure modifies large-scale eco-evolutionary patterns by changing the exponent. For $r=3.5$, a high proportion of individuals are highly social, leading to the formation of long-lived species with too-large population size, which can change the large-scale eco-evolutionary pattern by developing a mod in the tail. \textbf{C} and \textbf{D}: The power spectral density of the number of species for two different values of enhancement factors. For $r=3.5$, the evolution of highly social species leads to a cross-over in the power spectral density. The fitted exponents, $95$ percent confidence interval, and $R^2$ errors are shown in the legend. The effect of group size in a binary social perceptual public goods game. Parameter values: $\nu=10^{-3}$ and $N=10^4$}. 
	\label{Fig11}
\end{figure}

Two examples of population size distribution, for two values of enhancement factors, are shown in Figs. \ref{Fig11}\textbf{A} and \ref{Fig11}\textbf{B}, and reflect how the modification of large-scale eco-evolutionary patterns can range from a modification of the exponent, for $r=2.5$ to developing a peak for large population sizes for $r=3.5$. This is due to the high frequency of social individuals, forming highly social ``species'', whose sociality allows them to reach larger populations than that predicted by large-scale eco-evolutionary patterns. This evolution of social structure in turn, ``breaks'' this pattern, and leads to a faster decline of population size for small population sizes, which are predominantly individualistic species. Furthermore, investigation of the power spectral density of the number of species in Figs. \ref{Fig11}\textbf{C} and \ref{Fig11}\textbf{D} show that the evolution of sociality also modifies the spectral density of the number of species in the ecosystem, leading to a lower exponent in low frequencies, corresponding to large time-scales (which corresponds to highly social species who live longer and reach a higher population size than predicted by large-scale eco-evolutionary patterns), and a cross-over to a larger exponent (in absolute value) for larger frequencies.

\section{Discussion}
\label{Discussion}
Starting from a simple observation, that the behavior of living beings is determined by their perception, and perception, in turn, is an evolutionary construct, we have developed an evolutionary game theoretic picture of cooperative interactions, based on perceptual rationality. We have shown that perceptual diversity is an inevitable consequence of evolution in perceptually rational agents, which can drive social and behavioral diversity. While in a simple public goods game this perceptual diversity does not lead to behavioral diversity and individual decision-making remains consistent with predictions of pure rationality, under more complex evolutionary scenarios, a power law distributed diversity of rational decision-makers with consistent personalities is the outcome of evolution. 

While it has been suggested that bounded rationality is necessary to explain the allegedly irrational decision-making of individuals \cite{Simon1955}, this perspective suggests that individuals are always rational but from their own perspective. Instead of developing an objective perception of reality, individuals develop a distorted and diversified reflection of the outside world, which which we call ``reality'' and serves a functional role in maintaining the social structure by promoting cooperative behavior. Thus, embodying rational decision-making in the evolutionary process that shapes the realities of biological organisms, reveals that rational decision-making is the very root of the evolution of consistent personality differences and behavioral diversity that not only had been overlooked but had been framed as the exact opposite.

While this paper is a paper predominantly on evolutionary game theory, surprisingly, past work in evolutionary game theory has only a remote connection with this paper. Instead, as we discuss below, our work relates and interweaves several conceptual frameworks. 

\subsection{Connections with evolutionary game theory: conditional cooperation}

Much research has been devoted to the evolution of cooperation in public goods games under various modifications \cite{Szabo2007,Perc2008,Newton2018}. Of more relevance in the context of our paper is conditional cooperation. Within the bounded rationality framework, conditional strategies are typically pre-defined strategies: agents have prespecified decision-making rules and choose their strategies based on their conditional rules. While in the simplest applications, such as in games on networks, these behavioral rules are simple hard-wired rules \cite{Szabo2007}, in some cases, more flexible rules are allowed, for instance, by allowing agents to make decisions based on probability transition matrices, such as in signaling games \cite{Salahshour2019,Salahshour2020}, tag-based cooperation \cite{Riolo2001,Jansen2006}, or direct reciprocity \cite{Hilbe2018,Schmid2021}. 

Conceptually, conditional cooperation is a broad concept, and arguably, underlies many mechanisms for the evolution of cooperation, such as direct \cite{Trivers1971,Hilbe2018,Schmid2021} and indirect reciprocity \cite{Salahshour2022a,Nowak2005,Ohtsuki2006a,Boyd1989,Schmid2021,Nowak1998,Milinski2002,Kawakatsu2024}, tag-based mechanism \cite{Riolo2001,Jansen2006}, and coevolution of signaling and cooperation \cite{Salahshour2019,Salahshour2020}. In all these mechanisms, the ultimate cause of cooperation is the fact that agents withhold cooperation using conditional strategies. Most of the past work using conditional strategies has appealed to binary interactions. Although conditional cooperation has limited power in promoting cooperation in group interactions (due to the fact that it is not possible to selectively withhold cooperation from an individual, without affecting the whole group), examples of approaches applying this concept in public goods games are breaking ties and walking away from defectors \cite{Vukov}, conditional extortion \cite{Quan2019}, conditional contribution depending on the number of cooperators \cite{Szolnoki2012}, conditional strategies based on prestige \cite{Battu2020} or combined with heterogeneity \cite{Battu2018}, and conditional cooperation under multilevel selection \cite{Zhang2016}, and network heterogeneity in evolving network \cite{Li2017}.

In contrast to its typical use in bounded rationality, in the perceptual rationality framework we do not appeal to conditional cooperation directly (as a model assumption). However, diverse forms of conditional cooperation evolve in a self-organized manner due to the rationality of players. The self-organized emergence of conditional cooperation in our model has several consequences. Firstly, it contrasts the simplified way in which this concept is usually introduced in evolutionary games: as a model assumption and in the form of pre-defined, often hard-wired strategies. Secondly, it enters the model in a diversified form: the conditional cooperative strategies in our model can take diverse forms, that change their behavior in complex ways depending on the context. Thirdly, this provides a natural avenue to explain conditional cooperation observed in public goods experiments \cite{Keser2000,Fehr1999,Gintis2000}, which due to its diversity and complexity, better aligns behavioral experiments than the implementation of conditional cooperation in boundedly rational agents using pre-defined strategies. 

\subsection{Behavioral diversity and evolutionary game theory}

Another topic of importance in our paper is behavioral diversity. Increasing evidence has shown that, from humans to animals, behavioral diversity is an ineliminable face of decision-making  \cite{Sih2012,Reale2007,Sih2004}. In the context of cooperative interactions, behavioral diversity is linked to the prevalence of conditional cooperation \cite{Fehr1999,Gintis2000,Keser2000}. However, only recently attention to diversity has started to appear in evolutionary games. Nevertheless, diversity is typically introduced in evolutionary games only as a modeling assumption, for instance, in network connections \cite{Su2016,Pacheco2009,Santos2008,Perc2008,Santos2012,Zhu2014}, activity \cite{Van2009,Szolnoki2007}, individuals attributes \cite{Perc2008} \cite{Su2016,Van2009,Perc2008,Pacheco2009,Santos2008,Szolnoki2007,Perc2008,Santos2012}, production technology \cite{Shi2010,Hauser2019}, contributions \cite{Gao2010,Liu2017,Hauser2019}, and game strategies \cite{Zhang2010}. By doing so, these works have studied how different notions of diversity, introduced into models as behavioral rules, can affect evolutionary dynamics. Such approaches can shed some light on the functioning of diversity in strategic interactions and an important structural and evolutionary question remains unaddressed: why nature is diversity-rich and what organizing principles govern its evolution, structural, and functional role. While, we believe we are far from an understanding of this intriguing question, by showing that diversity can evolve in a self-organized way in a perceptual rationality framework, and due to the complexities inherent in rational decision-making, our framework can play a constructive role in addressing this question in the context of the evolution of behavior.

Besides, past works on the functional role of diversity often paint the picture of a world in which diversity can only improve cooperation. By contrast, empirical evidence has frequently suggested that different forms of diversity are often detrimental or, at best, neutral for cooperation \cite{Banerjee2005,Hjort2014,Dinesen2020,Baldwin2010,Habyarimana2007,Fung2014,Hjort2014}. There is a vast literature on the detrimental effect of ethnic diversity on cooperation \cite{Banerjee2005,Habyarimana2007,Baldwin2010,Hjort2014}. While this can result from many factors, from between-group inequality \cite{Baldwin2010} to strategy selection or technology mechanisms \cite{Habyarimana2007}, a consensus over why this phenomenon is widespread does not exist. We note that perceptual, or even social diversity, suggested by our model, can very reasonably be among the factors underlying this phenomenon.

Another interesting topic is the relation between diversity and power law distributions. Past works in evolutionary game theory have pre-defined diversity and measured it in qualitative terms \cite{Su2016,Pacheco2009,Santos2008,Santos2012,Zhu2014,Szolnoki2007,Van2009,Perc2008,Shi2010,Hauser2019,Gao2010,Liu2017}. While this can be a reasonable approach, research in statistical inference have established a mathematical notion of diversity by showing that power law distributions are distributions with maximum diversity \cite{Marsili2022,Marsili2013}. This mathematical foundation may raise the intriguing perspective of basing our understanding of social and behavioral diversity on a more mathematically rigorous notion of diversity. Besides, power law distributions are scale-invariant and can give rise to phenomena in all scales, leading to their appearance in many complex systems. This also includes diverse phenomena that govern social dynamics and behavior, e.g., \cite{Barabasi2005,Clauset2009}.

Given the ubiquity of power law distributions in many complex social and biological systems \cite{Marsili2022,Corominas2018,Newman2005,Brown2002,Stanley2000}, it is reasonable to expect evolutionary game theory should be able to establish at least some connection with such a notion of diversity. This notion of diversity and complex dynamics, however, have very rarely, and often marginally, appeared in evolutionary game theoretic models \cite{Sakiyama2021,Killingback1998,Zheng2007,Cherkashin2007,Sakamoto2023}. 

Our theory can address this gap by providing a framework that shows that such a notion of diversity can naturally evolve in a simple and scalable evolutionary game theoretic model. While some empirical evidence of power law distribution in human activities, such as power law distributed social networks \cite{Stephen2009,Muchnik2013,Clauset2009}, relate to some of our findings (noting that power-law degree distribution in such networks can be considered as a measure of social diversity \cite{Santos2012}), we are not aware of empirical examinations which are more directly related to some of our theoretical findings regarding power-law distributed behavioral, social, and perceptual diversity. In this regard, our work can encourage an empirical investigation of power-law distributed diversity in strategic interactions.

\subsection{Previous work on the evolution of rationality and perception}	

While evolutionary game theory is dominated by bounded rationality, some work within the evolution of preferences considers the evolution of rationality \cite{Newton2018,Alger2019,Konigstein2000,Samuelson2001,Guth1992}. By considering the evolution of preferences of rational agents, these works have shown how objective rationality can be evolutionarily unstable in different two-person two-strategy games, such as trust or ultimatum game \cite{Guth1998,Huck1999}, or generic two-person two-strategy games \cite{Dekel2007}, and lead to reciprocal behavior \cite{Guth1992}. Furthermore, existence theorems are provided according to which, a deviation from pure rationality can be generic \cite{Heifetz2007}. This body of literature demonstrates that objective rationality in game theory can be evolutionarily unstable under various conditions \cite{Newton2018,Alger2019}. However, as we have seen here, the conclusion that pure rationality is evolutionarily unstable, supported by many of these works in two-person games \cite{Guth1998,Heifetz2007,Konigstein2000,Dekel2007,Samuelson2001,Huck1999,Guth1992}, may not be straightforwardly generalizable to N-person games and in the absence of more complicated social structures (such as social information use, which was considered here). 

This body of literature on the evolution of rationality is also closely related to the evolution of perception, as in such indirect evolutionary approaches, while players are rational, their preferences, which arguably could be closely related to their perceptions are evolvable. However, a different line of work has more directly studied the evolution of perception. Using evolutionary game theoretic models, it has been studied how the evolution of perception can lead to the evolution of irrational strategies \cite{Hoffman2015}, deviations from reality \cite{Prakash2021}, or individual movement \cite{Swain2021}.

\subsection{The evolution of social information use}

Social information use is widespread in social animals, from insects to humans \cite{Evans2016, Laland2004,Kendal2018}, has attracted much attention \cite{Dall2005,Goodale2010,Mesoudi2016}, and is core to many opinion dynamics and collective information acquisition models. However, surprisingly, its implications and evolution in strategic decision-making do not seem to have been studied in evolutionary game theory. This fact seems to originate from the nature of the modeling approach used in evolutionary game theory, in which individuals are devoid of perception and are often rule-following agents. Consequently, the closest conceptual precedent to susceptibility to social information seems to be some forms of conditional cooperation, conditioning own's strategy on others' strategy, which was reviewed before. 

A relatively different way that researchers in evolutionary game theory have looked at social learning is via the very nature of evolutionary dynamics. In this regard, arguing not fitness, but social learning may govern evolution, some research has examined the effect of replacing the usual payoff comparison rule in updating the strategy (or combining it with) conformist social learning, according to which individuals opt for the most common strategy rather than the more successful one \cite{Molleman2013,Guzmán2007,Quan2022,Szolnoki2015,Szolnoki2012b}. Yet, another overlapping direction is implementing conformist learning in evolving networks by using such learning rules to decide who to play with \cite{Yang2020}. It is noteworthy to mention that this is usually examined in binary interactions (prisoner's dilemma) but not group interactions (public goods game).

We note that the evolution of social learning as a separate topic from the evolution of cooperation has been subject to some research to shed light on how individuals may use different strategies to integrate social and personal information, \cite{Kendal2009,Henrich1998,Morgan2012,Kendal2018,Mesoudi2016}, for instance, using Bayesian learning \cite{Perreault2012}, in the context of animal movement \cite{Guttal2010}, in game-theoretic contexts \cite{Schlag1998}, or conformist social learning \cite{Henrich1998}. 

In passing, we note that the evolution of sociality has many aspects, from solving diverse social dilemmas, such as cooperation in resource acquisition and parental care, to group formation and dispersal \cite{Garcia2013,Gadagkar1990,Alexander1974,Purcell2012}. While our work is tangential to this topic, we have focused on the evolution of cooperation and social information use, and thus, limit our discussion to these topics. 

\subsection{Rationality and personality}

Animal personality can be defined as between-individual variations which exhibit sufficient temporal stability and contexual consistency \cite{Kaiser2021}. The evolution of animal personalities over diverse dimensions of behavior, such as social information use \cite{Mesoudi2016}, has been subject to much research \cite{Dingemanse2010,Wolf2012}. While the diversity of the proposed models renders a classification of models a difficult task, many models show that adaptive personality differences can result from different forms of trade-offs, such as life-history trade-offs \cite{Wolf2008a}, growth trade-offs \cite{Biro2008}, or trade-offs resulting from social roles \cite{Rands2003}. Such trade-offs can result from density-dependent selection \cite{Wolf2008,Wolf2008a}, stochastic effects \cite{Rands2003}, or spatio-temporal environmental variations \cite{Stamps2007,Biro2008}. Furthermore, it is shown that these trade-offs can emerge naturally in eco-evolutionary dynamics based on energy use \cite{Salahshour2023,Salahshour2024}.

Other models have shown that variation within populations can result from different factors such as co-evolution of responsiveness and variation \cite{Dall2004,McNamara2009}, social niche by driving specialization and reducing conflict \cite{Bergmüller2010}, or co-evolution of different strategies exhibiting a synergistic relationship such as in cooperative \cite{Salahshour2021} or communicating strategies \cite{Botero2010}. This set of work reveals various mechanisms through which polymorphic populations composed of individuals with different roles may evolve. Besides, some of these works also satisfy more strict definitions of personality by showing that the evolved personalities can be consistent across contexts, for instance, due to life history trade-offs that render such contextual consistency adaptive \cite{Wolf2008a} 

Our work shows that rational decision-making can provide a natural road to the evolution of within-population behavioral variation, exhibiting contextual consistency in two different aspects of behavior, economic perceptions and social information use, which satisfies a rather strict notion of animal personality. This conclusion may seem counterintuitive given that historically rationality had been considered as the exact opposite of, and devoid of any diversity, giving rise to the fact that a shift away from pure rationality to bounded rationality had been perceived to be essential to account for behavioral diversity \cite{Simon1955}. Such a focus on behavior has, in turn, pervaded all the past theoretical work on the evolution of consistent personalities, leading to overlooking the very important role that rational decision-making, together with evolvable perceptions and social traits, may have played in driving consistent personality differences and behavioral diversity.

\subsection{Large-scale eco-evolutionary patterns}

Large-scale eco-evolutionary patterns are regularities, often in the form of power laws exhibited by diverse ecosystems \cite{Brown2002,Marquet2005}. While such patterns have been subject to research mostly in ecology and evolution, similar patterns have been found in economic and social systems \cite{Stanley2000}, paralleling the insights from our study that such patterns can be generalizable to power laws governing behavior within populations. Here, we have focused on a limited set of such patterns exhibited by our model, including fluctuation-scaling laws (Taylor's power law) \cite{Taylor1961,Taylor1984,Eisler2008}, species distribution \cite{McGill2007,Baldridge,Matthews2015}, species lifespan \cite{Drossel2001,Sole1997,Sole1998,Drossel1998,Pigolotti2005}, and the number of species in evolutionary times \cite{Drossel2001,Sole1996,Sole1997}.

While the generality of large-scale eco-evolutionary patterns may suggest they can serve as a testing ground for evolutionary game theoretic models, past studies in evolutionary games have not established a connection with large-scale eco-evolutionary patterns. Although in principle, likely, this has been also possible even in a usual bounded rationality framework, our framework readily provides insights into large-scale eco-evolutionary patterns by showing that large-scale eco-evolutionary patterns result from simple aspects of the evolutionary process, a stochastic multiplicative process, provided diversity is maintained in the population.

The generality of these patterns in a perceptual rationality framework allowed us to reach insights into how the evolution of social structure modifies large-scale eco-evolutionary patterns. In this regard, our findings link the microscopic social structure, i.e., diverse co-existing personality types exhibiting competition or complementation relationships, with emergent macroscopic patterns governing population dynamics. This insight can be a constructive step towards a better understanding of how new levels of social organization can evolve across the scales of organization and modify large-scale regularities in biological and social systems \cite{Levin1992}.

Finally, the evolutionary dynamics depicted here, on the one hand, admits an interpretation of an ecosystem as a public resource shared by diverse organisms, and on the other hand, is well consistent with the evolution of behavior within a species, where, ``species" can refer to behavioral types. This suggests similar large-scale eco-evolutionary patterns to those observed in between species, can be at work within species, governing the dynamics of social and behavioral diversity.

\subsection{Conclusion}

Our study suggests that accounting for the perceptual nature of decision-making can introduce a new perspective in evolutionary game theory with the potential to shed light on the evolution of diverse aspects of behavior. Here, we have demonstrated this point by introducing some modifications to the public goods game that can account for the evolvable perception of individuals. By introducing this perspective, we hope that our study encourages a better appreciation of aspects of decision-making in living organisms that have remained largely neglected.

\section{Materials and Methods}
\label{Methods}

\subsection{Statistics and reproducibility}
\label{Repro}
In the simulations presented in the main text, unless otherwise stated, we have set $N=10^4$ and $\nu=10^{-3}$. The effect of variations in mutation rate and population size are investigated in \red{S.5 and S.6}. The simulations in a structured population are performed in a community of agents residing on a first nearest-neighbor square lattice with periodic boundaries and von Neumann connectivity with linear dimension $L=100$. 

In Fig. \ref{Fig2}, the simulations in \textbf{A} are run for $5000$ time steps (except for perceptual public goods game in a well-mixed population, which is run for $7000$ time steps), and a time average after $T=4000$ is taken. In \textbf{B} and \textbf{C}, the simulations are run for $15000$ time steps, and the statistics are recorded after $T=10000$ in intervals of $10$ timesteps. A time average is taken over these statistics. In Fig. \ref{Fig3} and \ref{Fig4}, a sample of simulations run for $T=7000$ time steps is used. The time averages are taken after time $T=4000$. For the histogram in Fig. \ref{Fig3}\textbf{B}, a sample of $100$ simulations, run for $7000$, where the statistics are calculated after $T=5000$ is used. In Fig. \ref{Fig5}, a sample of $100$ simulations run for $15000$ timesteps is used. The statistics were recorded in intervals of $10$ time steps after time $10000$. 

In Figs. \ref{Fig6} and \ref{Fig7}, the simulations are run for $5000$ timesteps, and an average over the last $1000$ time steps is performed. A sample of $4$ simulations is used. The replicator dynamics is numerically solved for $5000$ timesteps. In Fig. \ref{Fig8}, the replicator dynamics is numerically solved for $10000$ time steps. For $g=3$, a time average over the last $5000$ time steps is taken (sampled each $5$ step). In Fig. \ref{Fig10}, the simulations are performed for $T=7000$ timesteps. A sample of $10$ simulations is used for each condition. In Fig. \ref{Fig11} the simulations are performed for $40000$ timesteps and the statistics are calculated after the first $2000$ timesteps.

All the simulations start with the worst initial conditions for the evolution of cooperation, in which all the individuals are asocial ($s=0$), have zero perception of public goods benefit, and uniformly at random distributed perceptions of public good cost. We have checked that, behaviorally, the dynamics do not show bistability. However, in the replicator dynamics of the social perceptual public goods game, perceptually, the dynamics can show multistability. That is, while the average behavior (cooperation) is the same starting from all the initial conditions, different distributions of perceptions consistent with the same behavioral outcome are possible starting with different initial conditions. This is shown in the \red{Supplementary Note. 8} using replicator dynamics. The replicator dynamics in Figs. \ref{Fig8} and \ref{Fig9} are solved starting from a homogeneous initial condition where the initial frequency of all the strategies is the same.

For behavioral power laws (cost and benefit perceptions and sociality) presented in the main text, we have used inverse quantities. An exponent $2-\tau$ for the inverse of a variable implies an exponent $\tau$ for that variable. Inverse variables are used to suppress the noise in histograms, since we have used logarithmically binned data, leading to high fluctuations for too-small bin sizes. For all the distributions the results of analysis for the original variables are presented in the \red{Supplementary Note. 4}.

``Species'' are defined as all the individuals with similar costs or benefits of public goods. Designating such populations as a species originates from the fact that all such individuals originate from a common ancestor and are perceptually and behaviorally similar. In a simple perceptual public goods game, we have based our analysis on all the individuals having the same perceived benefit, and in the social perceptual public goods game, we have based our analysis on individuals with the same perceived, cost, benefit, and sociality. The precise choice of the variable does not affect the results, as the important factor is that all the individuals within a species originate from a common ancestor, which is satisfied by all the choices. This observation can have important empirical consequences, by suggesting large-scale eco-evolutionary patterns also hold for sub-species, or sub-populations within a species, such as behavioral types. In this regard, our model suggests power laws similar to large-scale eco-evolutionary patterns may exist within species that govern the structure of behavior.

We note that in our framework individuals are behaviorally flexible, and the same individual may cooperate or defect in different contexts. Thus, the terms cooperating and defecting individuals should not be confused with the terms cooperators and defectors usually used in boundedly rational agents and are determined based on hard-wired strategies. Rather, these are defined based on the actions of the individuals. In the statistics that refer to cooperating and defecting individuals, we have based the calculations on the actions of individuals in an instance of the game. That is cooperation is defined as the counts of cooperation divided by the total of cooperation and defection in the population.

In The Supplementary Note. 3, we consider a modified version of the model in which individuals reproduce with a probability proportional to the exponential of their payoffs. With this choice, cooperation evolves in a simple perceptual public goods game played in a structured population. Nevertheless, perceptual power laws hold in such a model, as well.

\subsection{Replicator-mutator dynamics}
\label{RD}
The model in a well-mixed population can be solved in terms of the replicator-mutator dynamics \cite{Nowak2006b}. To do so, it is necessary to discretize the strategies (due to the challenge in calculating the Nash equilibrium). While the model can be discretized using any number of possible values for perceived cost, benefit, and sociality, we investigate a simplification to binary traits. This choice is because of the importance of binary trait scenario and its simplicity. This allows us to gain additional insights into the functional role of diversity in the evolutionary dynamics of cooperation.

We begin by deriving the replicator-mutator dynamics for the social perceptual public goods game. Discrete replicator dynamics reads as:
\begin{align}
	&	\rho_{cbs}(t+1)=\nonumber\\&\sum_{c'b's'}\nu_{cbs}^{c'b's'}\rho_{c'b's'}(t)\frac{\bar{\pi}_{c'b's'}(t)}{\sum_{c''b''s''}\rho_{c''b''s''}(t)\bar{\pi}_{c''b''s''}(t)}.
	\label{eqrepMM}
\end{align}
Here, $cbs$, $c'b's'$, and $c''b''s''$, refer to behavioral types, given by the value of the three traits, perception of the public good cost, $c\in\{0,1\}$, benefit, $b\in\{0,1\}$, and sociality, $s\in\{0,1\}$, respectively. $\rho_{cbs}$ is the frequency of type ${cbs}$, $\bar{\pi}_{cbs}$ is the expected payoff of type $cbs$, and $\nu_{{cbs}}^{c'b's'}$ is the mutation rate from the type $c'b's'$ to type $cbs$. Under our assumption that mutations in different traits occur independently, these can be written in terms of the probability of mutation $\nu$, as follows. for transitions that require no mutation we have, $\nu_{x}^{x}=(1 - 3\nu + 3\nu^2 - \nu^3)$. For those transitions that require one mutation, we have $\nu_{x}^{x'}=(1 - \nu)^2 \nu$. For those requiring two mutations, we have, $\nu_{x}^{x'}=(1 - \nu) \nu^2$, and for those requiring three mutations we have $\nu_{x}^{x'}=\nu^3$.

To use the replicator-mutator equation, eq. \eqref{eqrepMM}, we need expressions for the expected payoff of all the types. These are given by the following equations:
\begin{align}
	\bar{\pi}_x=\sum_{n_{000}+..+n_{111}=g-1} \frac{(g-1)!}{\prod_xn_x!}\prod_x{\rho_x}^{n_x}\pi(x,n_{000},..,n_{111}).
\end{align}
Where, the summation is over all the types subject to the condition, $n_{000}+..+n_{111}=g-1$. Here, $n_x$ is the number of groupmates, of type $x$, of a focal individual. The expression in front of summation is the binomial coefficient giving the probability of the group composition. Multiplying this with the payoff of the focal individual in such a group, $\pi(x,n_{000},..,n_{111})$, and summing over all the possible group compositions, gives the expected payoff of a focal individual with type $x$.

Now, the key term to be calculated is the expected payoff of a focal individual with type $x$ in a group with composition $n_{000}$,..,$n_{111}$. To do this, we need to calculate the Nash equilibrium of a public goods game, composed of $n_{000}$,..,$n_{111}$ individuals, of type, $000$,..,$111$, respectively. We first calculate the effective perceived cost and benefit for the individuals. For individualistic types, this is simply the perceived cost and benefit of the individuals. For collectivistic individuals, however, this is the average perceived cost and benefit of the individuals composing the group, including the focal individual. This is equal to $(n_{c=1}+\delta_{c_f,1})/g$, where $\delta_{c_f,1}$, is a delta function which is equal to one if the perceived cost of the focal individual is equal to $1$ and zero otherwise, and $n_{c=1}$ is the number of groupmates with perceived cost equal to $1$, which can be read based on the group composition, $n_{000}$,..,$n_{111}$. The effective perceived benefit of the collectivistic individuals can be calculated similarly.

Having the group composition, now it is possible to calculate the Nash equilibrium of the game as the configuration where no individual has a unilateral incentive to increase its payoff \cite{Nash1950}. That is, $(\sigma_i, \sigma_{-i})$, is a Nash equilibrium if and only if, for each individual i, $\pi_{\sigma_i, \sigma_{-i}}$ is not smaller than $\pi_{\sigma_i', \sigma_{-i}}$. Here, $\sigma_i\in{C,D}$ is the strategy of individual $i$ and $\sigma_{-i}=\{\sigma_j|j\neq i\}$ refers to the strategies of all the other individuals in the group. $C$ and $D$, are possible actions of the game, cooperation (investing) and defection (not investing). While the existence of Nash equilibrium is guaranteed \cite{Nash1950}, in some cases (when an individual has equal cost and benefit of the public good), it can happen that the game has more than one Nash equilibrium. If such cases happen, we calculate the average payoff of all the Nash equilibrium (while these case are rare in the continuous version of the model, they can be more common in the binary version).

The results presented in Figs. \ref{Fig6}, \ref{Fig7}, \ref{Fig8} and \ref{Fig9} show the numerical solutions of the replicator dynamics (often together with simulation results). The replicator dynamics for perceptual public goods games (in the absence of evolvable sociality), which is studied in the Supplementary Information, can be driven in a similar way. In this case, possible types are reduced to four, which can be shown by $cb$, where, $c,b\in\{0,1\}$.

%	\subsection{emprical data}

%	4471 species.

%\subsection{Relation to previous works}
%\label{Literature}
%Our work interweaves several conceptual frameworks. In this section, we provide a review of the relevant literature for our study.

\section{Supporting information}

\paragraph*{S1 Appendix.}
\label{S1_Appendix}
Supplementary analysis of the model supporting the findings of the manuscript is provided in the Supplemental Material Text.

\section{Acknowledgment} The author is thankful to Iain Couzin and Raghavendra Gadagkar for insightful comments. The author acknowledges funding from German Research Foundation (DFG – Deutsche Forschungsgemeinschaft) under Germany's Excellence Strategy - EXC 2117-422037984.

\nolinenumbers

% Either type in your references using
% \begin{thebibliography}{}
	% \bibitem{}
	% Text
	% \end{thebibliography}
%
% or
%
% Compile your BiBTeX database using our plos2015.bst
% style file and paste the contents of your .bbl file
% here. See http://journals.plos.org/plosone/s/latex for 
% step-by-step instructions.
% 

\newpage
\clearpage
\pagebreak



\begin{table*}[h!]
	\centering
	\begin{tabular}{|p{2.5cm}|c|c|}
		\hline
		\multicolumn{1}{|p{2.5cm}|}{\textbf{behavioral power-laws}} & \multicolumn{2}{c|}{\textbf{perceptual public goods game}} \\
		\hline
		& well-mixed (r = 2, 3, 4.7, 5) & structured (r = 2, 3, 4.7, 5)\\ 
		\hline
		benefit-to-cost ratio--- right tail &-2.06, -2.03, -2.04, -1.90&-2.03, -2.04, -2.33, -2.03 \\ 
		\hline
		benefit-to-cost ratio--- left tail&2.00, 1.93, 1.98, 1.91&1.98, 2.01, 2.03, 1.99\\
		\hline
		inverse perceived cost &-3.02, -2.81, -2.48, -1.93&-2.93, -2.77, -2.52, -2.12 \\ 
		\hline
		inverse perceived benefit&-1.96, -1.93, -2.08, -1.88&-1.83, -2.07, -2.07, -2.10\\
		\hline
		%			\multicolumn{3}{|c|}{\textbf{perceptual public goods game with evolvable sociality}} \\
		\multicolumn{1}{|p{2.5cm}|}{} & \multicolumn{2}{c|}{\textbf{social perceptual public goods game}} \\
		\hline
		& well-mixed (r = 1.5, 2.5, 3.5, 4.5) & structured (r = 1.5, 2.5, 3.5, 4.5)\\ 
		\hline
		benefit-to-cost ratio--- right tail&-1.98, -1.36, -1.32, -1.38&-2.03, -1.28, -1.41, -1.52\\
		\hline
		benefit-to-cost ratio--- left tail&2.03, 1.74, 1.63, 1.84&2.00, 1.33, 1.14, 1.69\\
		\hline
		inverse perceived cost&-2.39, -1.23, -1.21, -1.28&-1.99, -1.13, -1.25, -1.52\\
		\hline
		inverse perceived benefit&-1.98, -1.34, -1.23, -1.26&-1.97, -1.05, -1.28, -1.38\\
		\hline
		inverse sociality of social perceptions ($b>c$)&-2.13, -1.20, -1.47, -1.64&-2.07, -1.84, -1.98, -1.92\\
		\hline
		inverse sociality of anti-social perceptions ($b<c$)&-1.75, -1.11, -0.87, -1.04&-1.33, -1.49, -1.65, -1.45\\
		\hline
		inverse sociality (for all)&-1.75, -1.16, -0.77, -1.26&-1.34, -1.46, -1.73, -1.82\\
		\hline
		\multicolumn{1}{|p{2.5cm}|}{\textbf{eco-evolutionary patterns}} & \multicolumn{2}{|c|}{\textbf{perceptual public goods game}} \\
		\hline
		& well-mixed (r = 1.5, 2.5, 3.5, 4.5) & structured (r = 1.5, 2.5, 3.5, 4.5)\\ 
		\hline
		lifespan & -1.95, -1.95, -1.95, -1.93&-1.88,-1.84,-1.86,-1.83 \\ 
		\hline
		population size&-1.23, -1.23, -1.24, -1.28&-1.30, -1.29, -1.32, -1.32\\
		\hline
		mean population  & -2.49, -2.51, -2.54, -2.53&-2.49, -2.51, -2.47, -2.52 \\ 
		\hline
		total population  & -1.61, -1.62, -1.62, -1.60 &-1.59, -1.58, -1.59, -1.57\\ 
		\hline
		Taylor's exponent&1.02,1.01,1.01,1.01 &1.02,1.02,1.03,1.02\\
		\hline
		spectral density  & -1.12, -0.99, -1.15, -0.99&-1.16, -1.26, -1.33, -1.15 \\ 
		\hline
\end{tabular}

\end{table*}

\begin{table*}[h!]
	\centering
	\begin{tabular}{|p{2.5cm}|c|c|}		
		\hline
		\multicolumn{1}{|p{2.5cm}|}{} & \multicolumn{2}{|c|}{\textbf{social perceptual public goods game}} \\
		\hline
		& well-mixed (r = 1.5, 2.5, 3.5, 4.5) & structured (r = 1.5, 2.5, 3.5, 4.5)\\ 
		\hline
		lifespan &-1.93, -2.10, -2.23, -1.98& -1.82, -1.87, -2.09, -1.81 \\ 
		\hline
		population size&-1.27, -1.59, -2.95, -1.63&-1.41, -1.54, -1.85, -1.50\\
		\hline
		mean population &-2.62, -2.82, -2.89, -2.82 & -2.85, -2.71, -2.77, -2.75 \\ 
		\hline
		total population &-1.60, -1.70, -1.74, -1.81& -1.56, -1.59, -1.72, -1.57\\ 
		\hline
		Taylor's exponent&1.02,1.03,1.02,1.02&1.03,1.02,1.01,1.02 \\ 
		\hline
		spectral density of number of species & -1.36, -1.16, -0.60 (cross-over), -1.06& -1.11, -1.47, -1.10, -1.40 \\ 
		\hline
	\end{tabular}
	\caption{Behavioral power-laws and large-scale eco-evolutionary patterns. The best estimate of the values of the exponents for behavioral power laws and large-scale eco-evolutionary patterns for four values of enhancement factors indicated in the top row is shown. spectral density refers to the spectral density of the number of species, and mean and total population refers to the mean and total population size over a species' lifespan. See the Supplementary Notes, S.4 and S.10 for $95$ percent confidence interval and $R^2$ errors. }
	\label{Table1}
\end{table*}

\end{document}